\documentclass{article}

\usepackage{multirow}
\usepackage{arxiv}
\usepackage[utf8]{inputenc} 
\usepackage[T1]{fontenc}    
\usepackage{hyperref}       
\usepackage{url}            
\usepackage{booktabs}       
\usepackage{amsfonts}       
\usepackage{nicefrac}       
\usepackage{microtype}      
\usepackage{graphicx}
\usepackage{natbib}
\usepackage{doi}
\usepackage{subfig}
\usepackage{booktabs}
\usepackage{makecell}
\usepackage{array}
\usepackage{footmisc}
\usepackage{listings}
\usepackage{booktabs}
\usepackage{siunitx}
\usepackage{tikz}
\usepackage[T1]{fontenc}
\usepackage{tcolorbox}
\tcbuselibrary{most}
\tcbuselibrary{breakable}
\tcbuselibrary{listings}
\usepackage{inconsolata} 
\usepackage{soul}        
\usepackage{xcolor}
\usepackage{enumitem}    
\usepackage{textcomp}    
\usepackage{textpos}
\usepackage{hyperref}
\usepackage{xcolor}
\usepackage{fontawesome}

\definecolor{bgray}{HTML}{F8F9FA}
\definecolor{borderblue}{HTML}{4A90E2}
\definecolor{titlebg}{HTML}{343A40}
\definecolor{hlyellow}{HTML}{FFF3CD} 

\sethlcolor{hlyellow}

\title{Pre-review to Peer review: Pitfalls of Automating Reviews using Large Language Models}
\author{
    \textbf{Akhil Pandey Akella$^{\S\dagger*}$} \quad
    \textbf{Harish Varma Siravuri$^{\ddag}$} \quad
    \textbf{Shaurya Rohatgi$^{\natural}$}\\[0.5em]
    $^{\S}$\small AllSci Corp, Sunwater Capital \\
    $^{\dagger}$\small Kellogg School of Management, Northwestern University \\
    $^{\ddag}$\small Dept. of Computer Science, Northern Illinois University \\
    $^{\natural}$\small Institute of Foundation Models, MBZUAI \\[0.3em]
    \texttt{$^{\S}$\small aakella@allsci.com} \quad
    \texttt{$^{\ddag}$\small hsiravuri@niu.edu} \quad
    \texttt{$^{\natural}$\small shaurya.rohtagi@mbzuai.ac.ae} \\[1em]
    \href{https://github.com/akhilpandey95/LMRSD}{\faGithub~GitHub} \quad
    \href{https://huggingface.co/datasets/akhilpandey95/LMRSD}{~HuggingFace}
    \thanks{$^{\dagger\S}$ Work shared jointly between the listed affiliations.}
}

\date{}

\begin{document}
\maketitle
\begin{abstract}
Large Language Models are versatile general-task solvers, and their capabilities can truly assist people with scholarly peer review as \textit{pre-review} agents, if not as fully autonomous \textit{peer-review} agents. While incredibly beneficial, automating academic peer-review, as a concept, raises concerns surrounding safety, research integrity, and the validity of the academic peer-review process. The majority of the studies performing a systematic evaluation of frontier LLMs generating reviews across science disciplines miss the mark on addressing the alignment/misalignment of reviews along with the utility of LLM generated reviews when compared against publication outcomes such as \textbf{Citations}, \textbf{Hit-papers}, \textbf{Novelty}, and \textbf{Disruption}. This paper presents an experimental study in which we gathered ground-truth reviewer ratings from OpenReview and used various frontier open-weight LLMs to generate reviews of papers to gauge the safety and reliability of incorporating LLMs into the scientific review pipeline. Our findings demonstrate the utility of frontier open-weight LLMs as pre-review screening agents despite highlighting fundamental misalignment risks when deployed as autonomous reviewers. Our results show that all models exhibit weak correlation with human peer reviewers (0.15), with systematic overestimation bias of 3-5 points and uniformly high confidence scores (8.0-9.0/10) despite prediction errors. However, we also observed that LLM reviews correlate more strongly with post-publication metrics than with human scores, suggesting potential utility as pre-review screening tools. Our findings highlight the potential and address the pitfalls of automating peer reviews with language models. We open-sourced our dataset $D_{LMRSD}$ to help the research community expand the safety framework of automating scientific reviews.
\end{abstract}


\section{Introduction}
Existing peer-review structure relies on human reviewers to comprehend the manuscript, based on the ideas and concepts presented in the scientific paper, perform a literature review to assess novelty, comprehend the experimental methodology, make sense of the results, determine if the experimental results are valid, correct, and generalizable, and finally gauge the viability of the paper to give a score. While the scores neither imply nor guarantee post publication outcomes such as citations, hit-papers, etc., they do signal the ability of the authors to argue their merits and compel a decision of ``\textit{accept}" or ``\textit{reject}" from the human reviewer. The peer-review process is not universal, and human reviewers widely differ in their approaches across disciplines to accomplish various responsibilities associated with the peer-review process \citep{staudinger-etal-2024-analysis}, \citep{rogers2020whatcanwedo}. The scientific community has diverse opinions \citep{zhuang2025large} about the benefits and perils of automating peer reviews with the seemingly obvious objective of easing the burden on human reviewers and lifting the entry barrier for academic publishing. Despite the optimism, each opinion often recommends some degree of human steering or feedback to ensure the review process is robust and foolproof with new technology adoption. While \textit{trust}, \textit{safety}, and \textit{reliability} are implicit assumptions given to human reviewers and reviews, Large Language Models(LLMs) cannot be afforded that implicit trust and a systematic assessment of LLMs in generating reviews purely from the textual content can be a good start. Pairing this ability to critique and generate reviews of scientific papers with the proximity of the post-publication outcomes will help us observe and understand the potential downstream efficiency gains and pitfalls of automating peer reviews in science. 

\begin{figure}[!htb]
    \centering
    \subfloat[\centering Joint distribution of reviewer's paper rating vs reviewer confidence across papers]{{\includegraphics[width=0.45\textwidth]{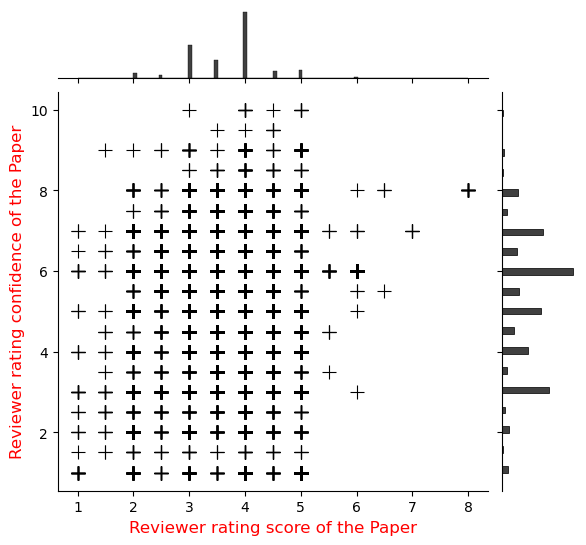} }}%
    \qquad
    \subfloat[\centering Open Review Paper Review Rating score distribution]{{\includegraphics[width=0.45\textwidth]{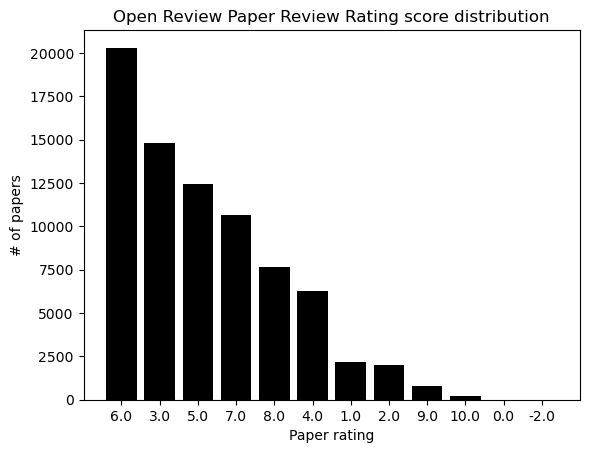} }}%
    \caption{$D_{LMRSD}$ data distribution (rating score, and confidence of reviews)}
    \label{fig:LMRSD_data_review}%
\end{figure}

In this paper, we address the following research questions surrounding automation of peer reviews using LLMs:
\begin{itemize}
    \item \textbf{R1:} How do frontier open-weight LLMs perform as peer reviewers compared to humans? \S\ref{RQ1}
    \item \textbf{R2:} Can we capture alignment/misalignment in LLM generated reviews by empirical methods? \S \ref{RQ2}
    \item \textbf{R3:} Can we assess usefulness of human-\textit{vs}-LLM reviews with publication outcomes? \S\ref{RQ3}
\end{itemize}

Given our research objectives, we wanted to gather a curated collection of scientific papers spread across various venues and disciplines in Science. Incidentally, there are several efforts on HuggingFace\footnote{https://huggingface.co/datasets?search=openreview} that gather and accommodate an aggregate dataset of papers from \textbf{OpenReview}\footnote{https://openreview.net/} platform. Several entries from the above HuggingFace URI included scientific works and datasets mostly associated with Computational Sciences, and most of these data sources include cleaned and post-processed information about the paper, authors, and reviewers. However, we observed that in most of these datasets, there was minimal to no information about impact metrics for publications. In addition, we barely noticed disambiguated linkages to publication IDs, author IDS and other relevant information that could help us gauge post-publication outcomes effectively. Gathering this information is crucial for evaluating peer reviews generated using LLMs. To this extent, to capture and \textbf{L}earn \textbf{M}eaningful \textbf{R}ewards about reviews of \textbf{S}cientific \textbf{D}ocuments, our paper introduces a unique dataset \textbf{$D_{LMRSD}$}.

Our work utilizes $D_{LMRSD}$ to answer the aforementioned research questions by: a.) Running Zero-shot inference using different open-weight dense and reasoning language models on the entire dataset to capture the comprehensive review distribution across papers. We subdivided the task of review generation into: i.) Generating idea-reviews for a paper with just the abstract, and ii.) Generating idea-reviews and peer-reviews with full-text. b.) Measure hypothetical review \textit{alignment/misalignment} using a visual framework, and c.) Post-publication outcome analysis on a subset of high-impact papers from $D_{LMRSD}$ to validate the usefulness of LLM reviews in identifying \textit{Novel} \cite{Uzzi2013AtypicalCA}, \textit{Disruptive} \cite{Wu2019LargeTD} and \textit{Hit papers} \cite{lin2023sciscinet} regardless of their purported agreement with ground-truth human reviews.

Our study can serve as a guiding framework for assessing the safety of using LLMs in automating peer reviews. Our findings show that most frontier open-weight language models exhibit strong overconfidence and weak performance compared to human ground-truth baselines. We noticed that most models are erroneous in evaluating papers with lower peer review scores but mostly do a decent job evaluating high impact papers. Additionally, we show most models exhibit weak alignment when we visually analyze the self-reported confidence \citep{xiong2023can} measured against predicted bias between LLM generated review scores and human peer review scores. Our results overwhelmingly show models inflate review scores by 3-5 points above human reviewers (LLM means: 7.5-9.0 vs. human medians: 3-7), while maintaining uniformly high confidence scores (8.0-9.0/10) regardless of prediction bias. Most intriguingly, we find that LLM reviews demonstrate stronger correlations with post-publication metrics such as $C_5$ (citation counts until 5 years), Novelty, and Conventionality than with human reviewer scores, suggesting that LLMs may capture different signals of scientific merit than human reviewers. These findings suggest that while LLMs could serve as useful pre-review screening tools for identifying conventionally impactful work, their current deployment as autonomous peer reviewers poses substantial risks due to systematic biases, overconfidence, and fundamental misalignment with human judgment particularly in identifying truly disruptive and novel scientific contributions.

We acknowledge the technical limitations of using LLMs to generate peer reviews of scientific works solely based on their textual content. However, our objectives with this study are localized to providing empirical evidence and offering an experimental framework to assess the viability of utilizing LLMs as pre-review agents, if not for peer reviews. Our efforts in this study are directed towards gaining insights into the benefits, risks, and pitfalls of automating peer reviews with LLMs, and towards developing a meaningful pathway to validate the generated reviews through post-publication outcomes rather than replacing the gold standard of human effort and judgment.

\section{Dataset}
OpenReview\footnote{https://openreview.net/} is a scholarly resource repository that publishes peer reviews for various scholarly papers and acts as a key source for transparent and accessible peer reviews. Ideally, any dataset extracted from openreview provides text, rating, and confidence of the review in a semi-structured format. While the papers and authors are disambiguated, the identities of the reviewers are still anonymized. We use a combination of \cite{hopner2025automatic}, and \cite{staudinger-etal-2024-analysis} to form a peer-review dataset $D_{LMRSD}$ comprising $77845$ paper reviews for all of our experiments. $D_{LMRSD}$ consists of $28033$ unique papers and the composition of venues for each of the unique papers can be observed from Table \ref{tab:venue_composition}.

Empirically, reviewer confidence in $D_{LMRSD}$ from Fig \ref{fig:LMRSD_data_review} a.) shows that reviewer confidence tends to be highest ($7-10$) when papers receive ratings in the range of $3-6$. While we cannot establish a causal link for this existence, we can clearly notice less reviewer confidence at the extremes, both for papers rated in the range $1-2$ and $8-10$. The marginal distribution from Fig \ref{fig:LMRSD_data_review} b.) reveals a strong negative skew in paper ratings, with the mode at 6.0 and a substantial number of papers receiving ratings of 3-7, while very few papers achieve top ratings of 8-10. This distribution mirrors the power law patterns observed in science of science literature\citep{fortunato2018science, brzezinski2015power}, where high-impact research follows a heavy-tailed distribution similar to how citation counts, breakthrough discoveries, and influential papers are concentrated among a small fraction of publications, while the majority cluster around average impact levels.

While $D_{LMRSD}$ is sufficient to evaluate the performance of LLM generated reviews against human scores, a true measure of real-world impacts for the paper reviews can arise from post-publication outcomes in science of science metrics such as \textbf{Citations}, \textbf{Novelty \& Conventionality Score}\citep{uzzi2013atypical}, \textbf{Disruptive Index}\citep{wu2019large}, \textbf{Hit-papers}\citep{lin2023sciscinet}. These are more meaningful and stand as empirical evidence of value for our study. For this reason we chose to integrate $D_{LMRSD}$ with SciSciNet \cite{lin2023sciscinet} since it helps us avoid recalculating some challenging pre-calculated metrics such as Citation, Level-0, and Level-1 Microsoft Academic Graph \cite{sinha2015overview} field information, and lastly information on papers that went on to become Hit-papers in the top 1\%, 5\% or 10\% in the respective fields. Each of these terms are elaborately described in SciSciNet's paper\cite{lin2023sciscinet}. 

The paper-IDs in $D_{LMRSD}$ did not include DOI, or other identifiers we could use to connect our dataset with SciSciNet easily. We therefore attempted to establish a cross-walk using paper title. This cross-walk resulted in a 100 percent match, thereby enriching the review information in $D_{LMRSD}$ with the metrics and outcomes information from SciSciNet. Although we didn't perform a validation of the linkage, surface-level inductive checks on 50 random samples gave us confidence in the cross-walk and linkage.

\section{Related Work}

Existing literature on automatically generating peer reviews primarily focuses on building peer-review datasets to gather qualitative information about the peer‑review process. \citet{lin2023moprd} introduces a multidisciplinary open review corpus covering the full review pipeline—paper metadata, version history, multiple reviewer comments, meta‑reviews, rebuttals, and editorial decisions. \citet{kang-etal-2018-dataset} introduce PeerRead, one of the first large public peer-review datasets, have supported both acceptance prediction and review modeling tasks. While \citet{szumega2023open} rely on OpenReview website to gather and extract \textit{accept}, \textit{reject} and other metadata about peer-review decisions.

Another area where our study intersects with existing literature is when LLMs are used to automate peer review. Yu et al. \cite{yu-etal-2024-automated} suggest a mismatch score and a loop to correct generated reviews for higher alignment. More broadly, recent surveys like \citet{zhuang2025large} on LLMs for automated scholarly review outline open challenges and trade-offs when deploying these models in the review pipeline. Although we didn't fine-tune a model on our dataset like \citet{idahlahmadi2025openreviewe}, we do see a pathway for researchers to utilize $D_{LMRSD}$. Lastly, to establish a systematic benchmark for peer-review datasets and help train models, we must need exhaustive tasks that are unique and comprehensive. The task ontology from \cite{staudinger-etal-2024-analysis} can help training LLMs on sub-tasks possible.

While these works are relevant and related, we evaluate whether generated reviews preserve reviewer intent, flag misalignment, and examine whether LLM generated feedback could distort acceptance outcomes. Unlike earlier works focused only on the review generation or decision prediction stage, we correlate LLM generated review ratings with downstream success indicators—such as citations, “hit‑paper” status, novelty and disruption scores offering an empirical testbed for alignment-to-impact.

\section{Methodology}

\begin{figure}[!htb]
    \centering
    \subfloat[\centering Token distribution across model families for prompt using Title, Abstract, and Keywords]{{\includegraphics[width=0.45\textwidth]{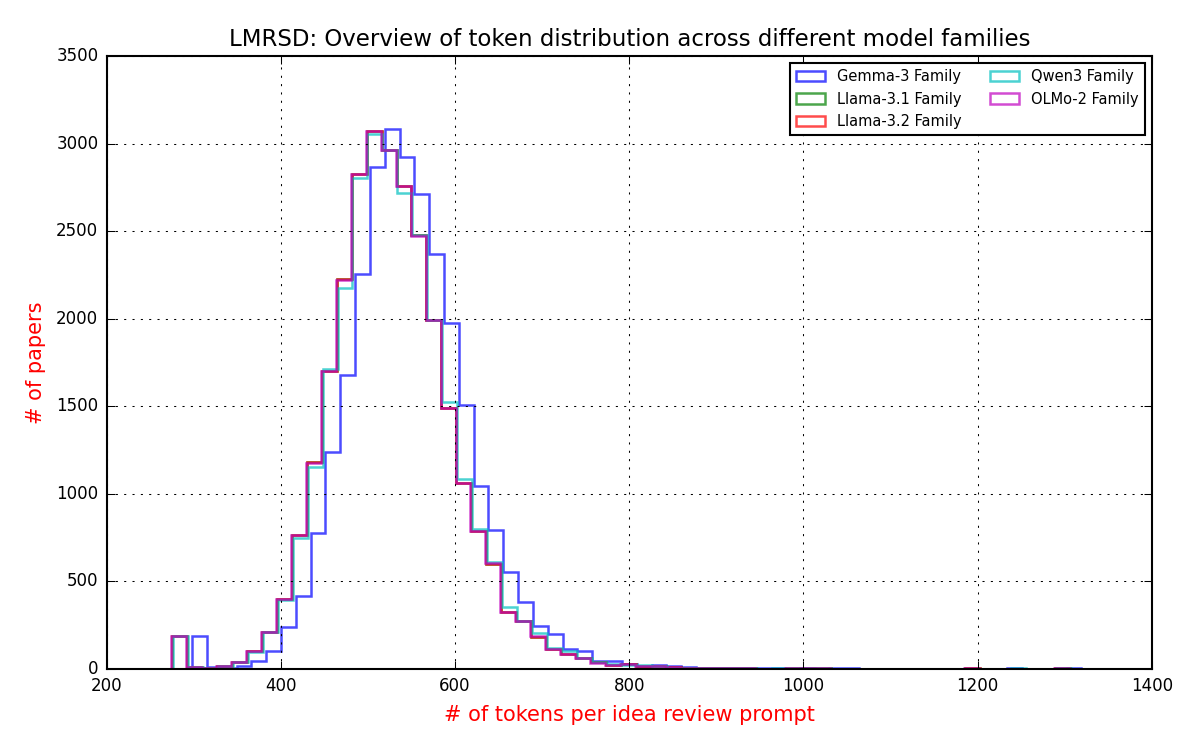} }}%
    \qquad
    \subfloat[\centering Token distribution across model families for prompt using Title, Keywords, and Full text.]{{\includegraphics[width=0.45\textwidth]{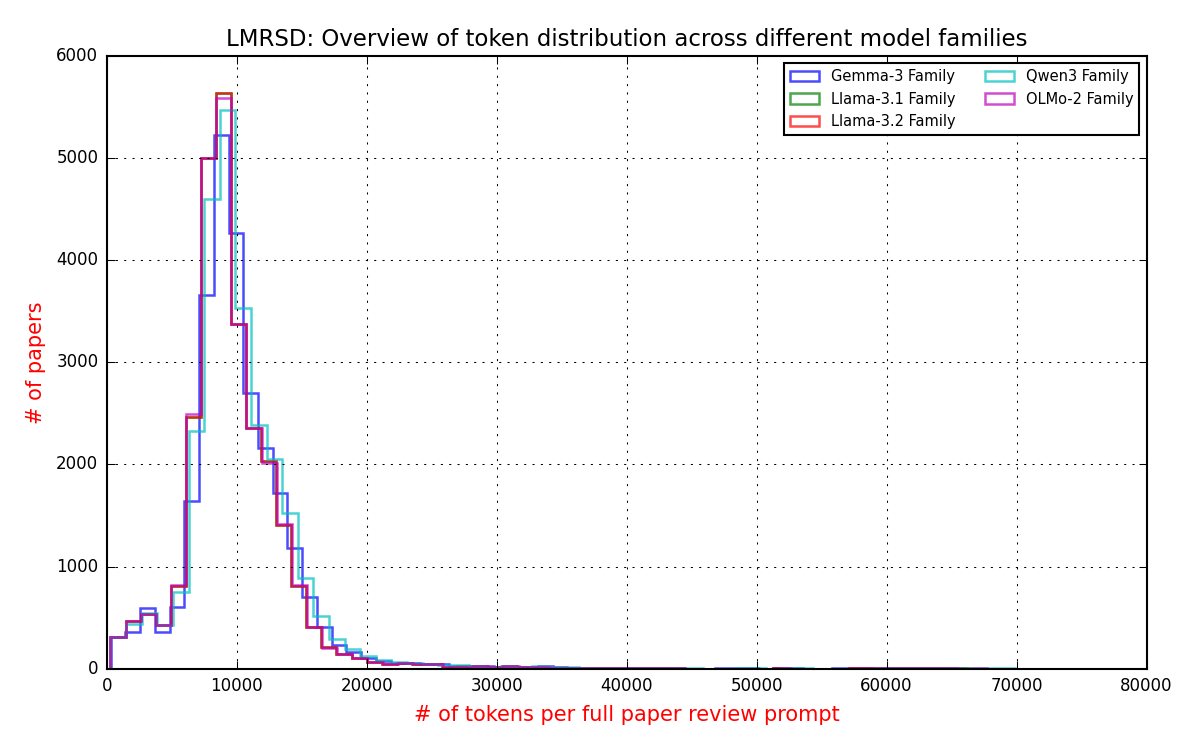} }}%

    \caption{Token distribution across model families for the prompts evaluating scientific manuscripts}
    \label{fig:data_overlay_hist}
\end{figure}

To evaluate the viability of open-weight LLMs in scientific peer review pipelines, we systematically assessed nine frontier open-weight models across two experiments: abstract-only idea evaluation (Experiment 1) and full-text manuscript review (Experiment 2). We report correlation metrics to measure alignment with human reviewers and error metrics to quantify prediction accuracy. We designed the evaluation framework aligned with the aforementioned research agenda to assess the capability (\ref{RQ1}), quality (\ref{RQ2}), and usefulness (\ref{RQ3}) of LLM generated reviews in their role as scientific peer reviewers. Our methodology demanded open-weight models chosen for their ability to process long contexts (>32k tokens) given the prompts (Appendix \ref{zero_shot_prompts}, Appendix \ref{zero_shot_prompts_idea_review}, Appendix \ref{zero_shot_ablations_prompts}) token distribution (Fig. \ref{fig:data_overlay_hist}) when full-texts/abstracts are provided as inputs. We included \textbf{Dense} models: \textit{Llama-3.3-70B-Instruct}, \textit{Llama-3.1-Tulu-3-70B}, \textit{Llama-3.3-Nemotron-Super-49B-v1.5}, \textit{Qwen3-32B}, and \textit{Gemma3-27b-it}; and \textbf{Reasoning} models: \textit{DeepSeek-R1-Distill-Llama-70B}, \textit{Qwen3-Next-80B-A3B-Thinking}, \textit{gpt-oss-20b}, and \textit{gpt-oss-120b} in our experiments. All models were deployed using vLLM or SGLang backends with temperature set to $0$, cumulative probability set to $1$, and seed set to $2025$ for reproducible outputs. We used three variations of $D_{LMRSD}$ dataset each catering to specific parts of our experimental study. These include:
\begin{itemize}
    \item \textbf{LMRSD.a:} Used for assessing and reviewing just the ideas in a scientific article using the paper title, keywords, and abstract.
    \item \textbf{LMRSD.b:} Used for assessing and reviewing the ideas and complete manuscript peer-review of the scientific article using paper title, keywords, and full-text.
    \item \textbf{LMRSD.c:} Same dataset as \textbf{LMRSD.b} filtered down by availability of C5, Hit paper (1\%, 5\%, and 10\%), Novelty, and Disruptive index metrics.
\end{itemize}

All scientific papers come with existing human peer review scores, including median and average reviewer ratings on a 1-10 scale. Each paper includes metadata such as title, abstract, keywords, and full text, along with post-publication impact metrics including citation counts (C3, C5, C10), disruption scores, and novelty measures (Atyp\_Median\_Z). For Experiment 4, we specifically curated a subset of papers categorized as "hit papers" based on their post-publication success (top 1\%, 5\%, and 10\% by citations). Further implementation details can be found in our code repository \footnote{https://github.com/akhilpandey95/LMRSD}.

\subsection{Frontier open-weight LLMs performance as peer-reviewers}
\label{RQ1}

Our preliminary objective was to assess whether LLMs can evaluate the core scientific ideas in a paper solely from its abstract, title, and keywords. This mirrors scenarios where rapid screening of submissions is needed. We used the prompt structure detailed in Appendix~\ref{zero_shot_prompts_idea_review} to evaluate five dense models on the \textbf{LMRSD.a} dataset ($n = 26,391$ papers). Since OpenReview does not provide separate idea-quality ratings, we treat the full-paper review scores as our reference ($y_{\text{true}}$) and compare against LLM generated idea scores ($\hat{y}_{\text{LLM}}$). Table~\ref{tab:exp1_corr_metrics} reveals uniformly low correlations across all models. The highest correlation amongst all, Gemma-3-27B-IT, achieves only $\rho = 0.153$, explaining less than 2.5\% of variance in human scores. Notably, Llama-3.3-70B-Instruct shows near-zero correlation despite being the largest dense model tested, suggesting that model scale alone does not predict alignment with human judgment.    

\begin{table}[htbp]
\centering
\caption{Correlation Metrics for Idea review | $\hat{y}_{LLM}$ vs $y_{true}$ paper review}
\label{tab:exp1_corr_metrics}
\begin{tabular}{lccc}
\toprule
\textbf{Model} & {$Pearson$} & {$Spearman$} & {$Kendall$} \\
\midrule
$\hat{y}_{LLM=Qwen3-32B}$ & 0.13262 & 0.131515 & 0.108729 \\
$\hat{y}_{LLM=Llama-3.3-70B-Instruct}$ & 0.009031 & 0.001268 & 0.000755 \\
$\hat{y}_{LLM=gemma-3-27b-it}$ & 0.151437 & 0.153084 & 0.128639 \\
$\hat{y}_{LLM=Llama-3.3-Nemotron-Super-49B}$ & 0.151439 & 0.1478 & 0.126441 \\
$\hat{y}_{LLM=Llama-3.1-Tulu-3-70B}$ & 0.12727 & 0.122341 & 0.105495 \\
\bottomrule
\end{tabular}
\end{table}

To answer \ref{RQ1}, we reported Mean Absolute Error (MAE), Root Mean Squared Error (RMSE), and Maximum Error to assess the performance of LLMs in generating peer review scores for scientific articles. The error metrics in Table~\ref{tab:exp1_metrics} indicate prediction deviations, and if we were to interpret the MAE values across all models, a range of 2.73 - 3.50 on a 10-point scale typically indicates that predicted peer review scores deviate by nearly 3 points from human scores. So, there is limited consensus between models and humans in which, the downstream effect of using the best model Qwen3-32B would mean a difference between \textit{reject} and  \textit{weak accept}. For a dataset this size \textbf{LMRSD.a} it is expected to have the maximum error (Max Error) reach 9.5 points, simply because there will be papers that the model would roll out a review score vastly different from the ground truth. Interestingly, Qwen3-32B achieves the lowest MAE despite not having the highest correlation, suggesting these metrics capture complementary aspects of performance. A natural question here is whether providing complete manuscripts improves LLM review quality.

\begin{table}[htbp]
\centering
\caption{Error Metrics for Idea review | $\hat{y}_{LLM}$ vs $y_{true}$ paper review}
\label{tab:exp1_metrics}
\begin{tabular}{lccc}
\toprule
\textbf{Model} & {\textbf{MAE}} & {\textbf{RMSE}} & {\textbf{Max Error}} \\
\midrule
Qwen3-32B & \textbf{2.731087} & \textbf{3.175855} & 8.5 \\
Llama-3.3-70B-Instruct & 2.836118 & 3.296270 & 8.5 \\
Gemma-3-27B-IT & 2.900707 & 3.319098 & 8.0 \\
Llama-3.3-Nemotron-Super-49B-v1.5 & 3.262146 & 3.647522 & 9.0 \\
Llama-3.1-Tulu-3-70B & 3.498581 & 3.864499 & 9.5 \\
\bottomrule
\end{tabular}
\end{table}

\begin{figure}[!htb]
    \centering
    {\includegraphics[width=0.75\linewidth]{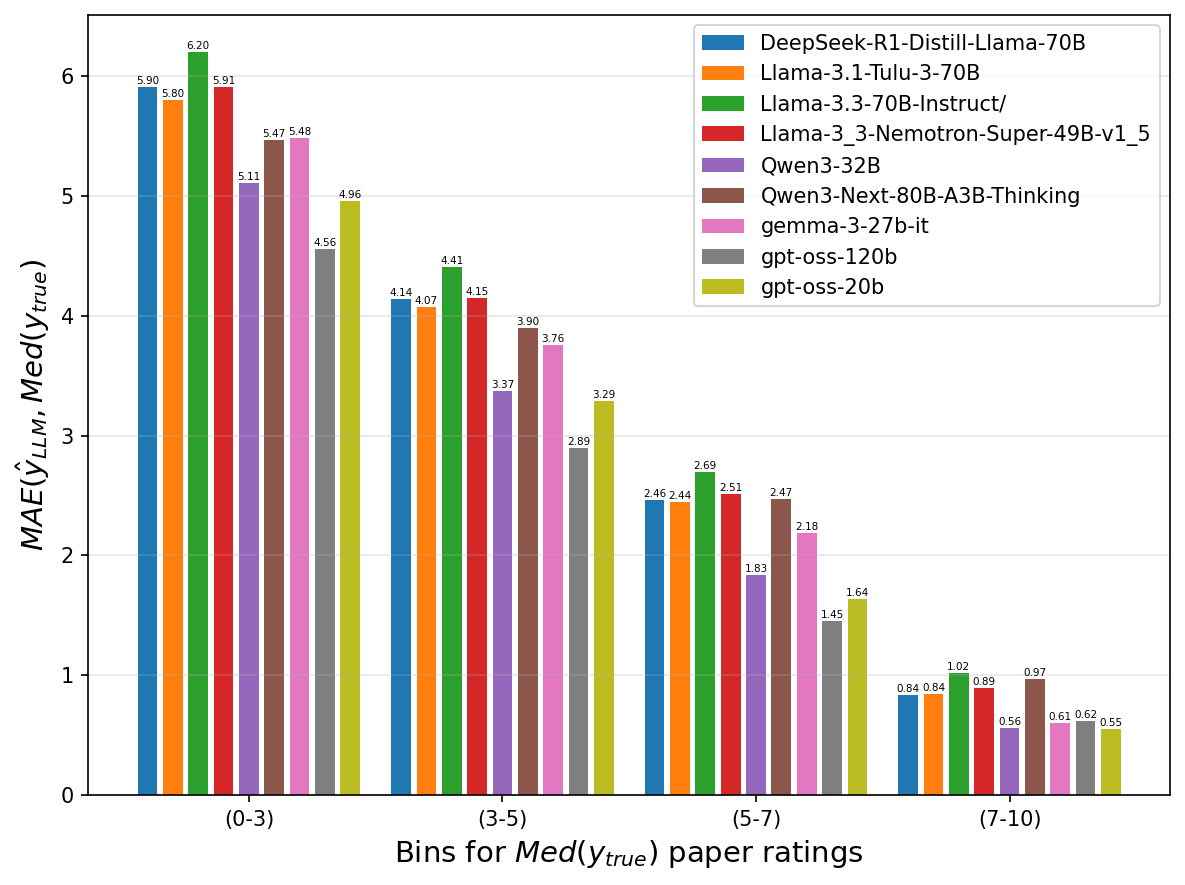} }
    \caption{Mean absolute error across different true score ranges for all models}
    \label{fig:lmrsd_mae}%
\end{figure}

We designed Experiment 2 to test this by providing full text to nine models: five dense architectures and four reasoning models capable of generating chain-of-thought traces (see Appendix~\ref{zero_shot_prompts} for prompts). Table~\ref{tab:exp2_metrics} reveals some interesting yet similar findings. Most evidently, despite access to complete manuscripts, reasoning models do not consistently outperform dense models. While gpt-oss-120B achieves the lowest MAE, DeepSeek-R1-Distill-70B received the highest MAE indicating limited consensus with human scores. Almost all of the models exhibit overly positive peer review scores, with mean predicted scores ranging from 7.56 to 8.98 on a 10-point scale. The above methodology provided us with structure and methodological rigor in evaluating frontier open-weight language models both as idea reviewers and peer-reviewers for scientific texts. Critically, we observe that despite having substantially more information (full paper text vs. abstract only), models do not show a proportionate improvement in peer review quality when compared against ground truth. Capturing the overall MAE for each model can squash and diminish problematic patterns across different review score zones for each paper. So we stratified the ground-truth score from $1-10$ into 4 bins and plotted the MAE for the predicted review score and median of all ground truth peer review scores for each model mentioned in Table \ref{tab:exp2_metrics}. Figure~\ref{fig:lmrsd_mae} shows MAE across four bins of human review scores, highlighting asymmetries across each range. All models perform worse on papers rated 0--3 by humans (MAE $\approx 4.5$--$6.2$) but achieve reasonable accuracy on papers rated 7--10 (MAE $\approx 0.55$--$1.02$). This translates to a \textbf{10-fold difference} in error magnitude between the lowest and highest scored papers. Clearly, this pattern is consistent across all nine models, regardless of architecture or reasoning capability. While it is difficult to understand the causality behind this asymmetry, a simple empirical observation here could be that when encountering papers that humans judge as fundamentally flawed, LLMs consistently assign scores near the middle or upper portion of the scale. Model sycophancy appears and extreme agreement is making LLMs unable or unwilling to render lower peer review scores, instead defaulting to inflated scores regardless of manuscript quality. 

When we combine this finding with the overestimation bias in the above confidence plots, we can observe that most frontier open-weight LLMs perform well as peer reviewers purely using the full-texts without any tool calls given the papers are rated highly by humans. From a deployment perspective, this simply means using LLMs to fully automate reviews, human reviewers or systems would have to be extra careful in relying on the judgement because they can systematically allow weak papers to pass initial screening while providing limited discriminative value.

\begin{table}[htbp]
\centering
\caption{Error Metrics for Experiment 2: Full-Text Review Performance}
\label{tab:exp2_metrics}
\begin{tabular}{lccc}
\toprule
\textbf{Model} & {\textbf{MAE}} & {\textbf{RMSE}} & {\textbf{Max Error}} \\
\midrule
GPT-OSS-120B & \textbf{2.127616} & \textbf{2.594593} & 8.0 \\
GPT-OSS-20B & 2.404172 & 2.865529 & 8.0 \\
Qwen3-32B & 2.484509 & 2.900157 & 8.0 \\
Gemma-3-27B-IT & 2.817559 & 3.184826 & 8.0 \\
Qwen3-Next-80B-A3B-Thinking & 3.062358 & 3.407408 & 9.0 \\
Llama-3.1-Tulu-3-70B & 3.087241 & 3.430558 & 9.0 \\
DeepSeek-R1-Distill-Llama-70B & 3.132514 & 3.486439 & 9.0 \\
Llama-3.3-Nemotron-Super-49B-v1.5 & 3.171045 & 3.514970 & 9.0 \\
Llama-3.3-70B-Instruct & 3.375088 & 3.692477 & 9.0 \\
\bottomrule
\end{tabular}
\end{table}

\subsection{Alignment/Misalignment of LLM generated reviews peer-reviewers}
\label{RQ2}

\textit{Alignment} as a premise is very popular in \textit{RLHF} and reinforcement learning in general, in that we are expressly attempting to imitate learning from human feedback using a targeted input-output data distribution. In the context of evaluating LLMs doing In-context learning for peer reviews, a fundamental question is whether their judgments exhibit meaningful alignment with human consensus, not merely in accuracy/error metrics, but in the structure of their uncertainty. Human reviewers, as shown in Fig \ref{fig:LMRSD_data_review}, display a joint distribution across rating scores (1 - 8) and self-reported confidence (1 - 10) that reflects genuine variation. Reviewers express lower confidence when papers fall outside their expertise, when methodological aspects are ambiguous, or when the contribution is difficult to assess. Its reasonable to expect an LLM doing peer review to exhibit similar behavior mimicking confidence that correlates with prediction accuracy. 

\begin{figure}[!htb]
    \centering
    \subfloat[\centering Llama-3.3-70B-Instruct]{{\includegraphics[width=0.47\textwidth]{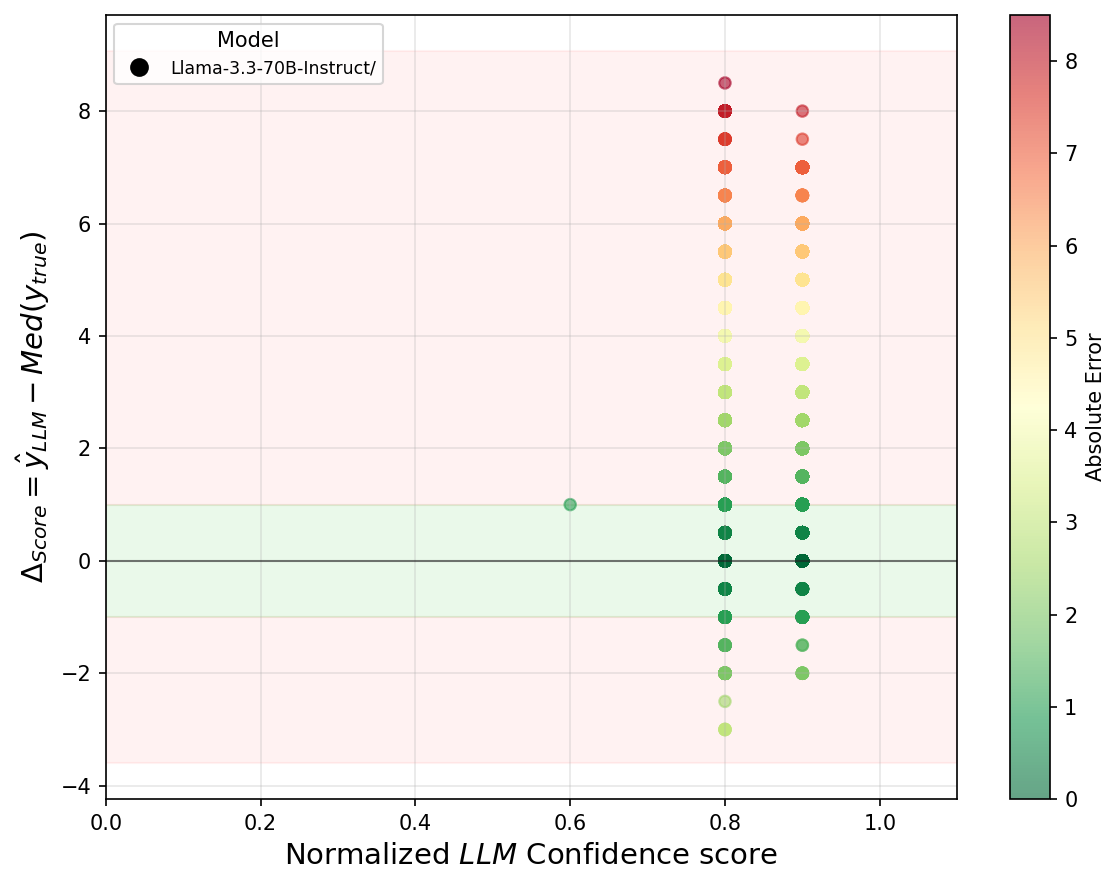} }}%
    \qquad
    \subfloat[\centering Qwen3-32B]{{\includegraphics[width=0.47\textwidth]{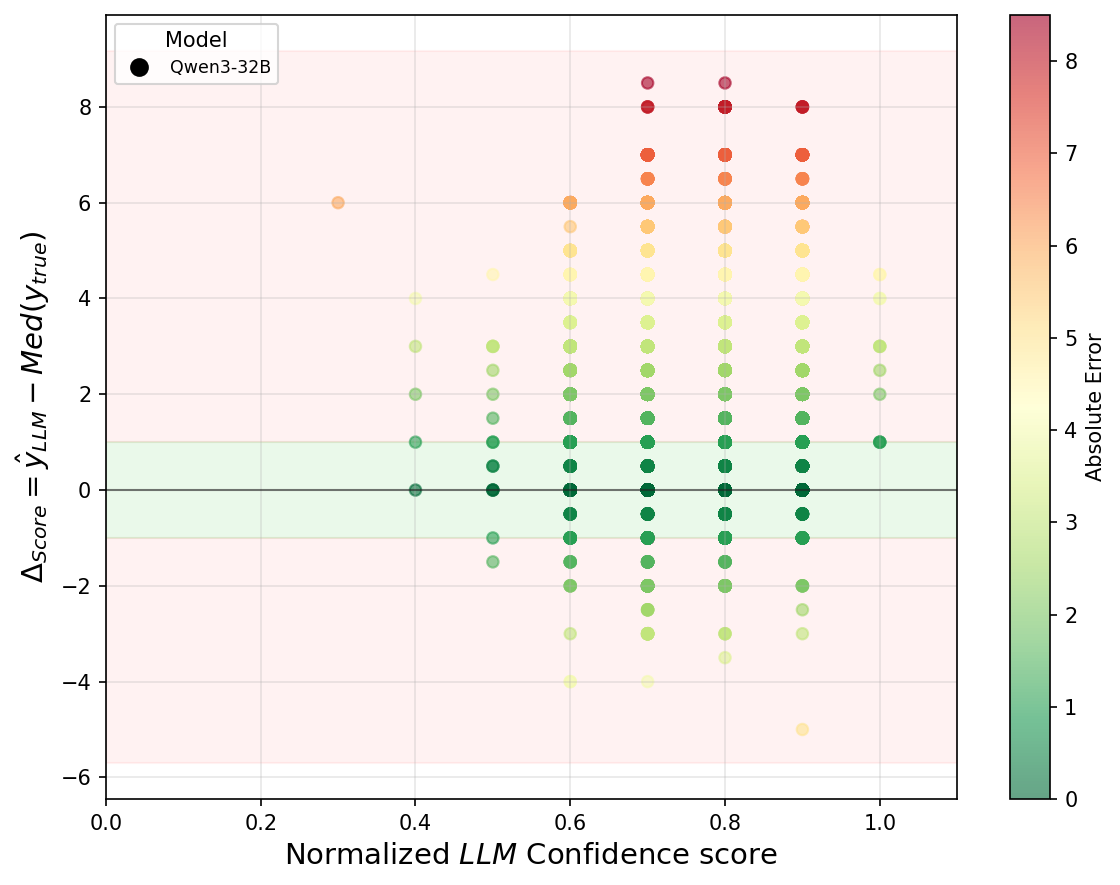} }}%
    \caption{Confidence plot when LLMs review ideas in a given paper using abstract}
    \label{fig:LMRSD_exp1_conf_plot}
\end{figure}

To probe this premise, we used a simple visualization that plots prediction bias ($\Delta_{\text{Score}} = \hat{y}_{\text{LLM}} - \text{Med}(y_{\text{true}})$) against the self-reported confidence score (normalized to $[0,1]$) for each model. We define an acceptable prediction zone as $\Delta_{\text{Score}} \in [-1, 1]$ (shown in green), corresponding to the typical inter-reviewer disagreement observed in peer review; prior work on OpenReview data reports mean absolute deviation between reviewers of approximately 1.0 - 1.5 points on a 10-point scale \citep{rogers2020whatcanwedo}, making this threshold a reasonable proxy for human level consistency. For a model to exhibit human like behavior, we would expect two observations from the language models: a.) predictions should cluster within this green zone when a model reports high confidence, and b.) lower confidence should accompany larger deviations from human consensus, reflecting appropriate uncertainty when judgments are unreliable. The confidence scores analyzed here are \textit{self-reported} confidence values generated by the LLM in response to our prompt requesting a confidence rating. These should not be conflated with predictive entropy or token-level probabilities of the model. Verbalized confidence reflects whatever heuristics the model has learned to associate with the word "confidence" during instruction tuning, which may or may not correlate with actual predictive reliability. Our analysis thus tests whether LLMs have learned to express calibrated uncertainty in peer review contexts, not whether they possess it internally.

\begin{figure}[!htb]
    \centering
    \qquad
    \subfloat[\centering DeepSeek-R1-Distill-Llama-70B]{{\includegraphics[width=0.45\linewidth]{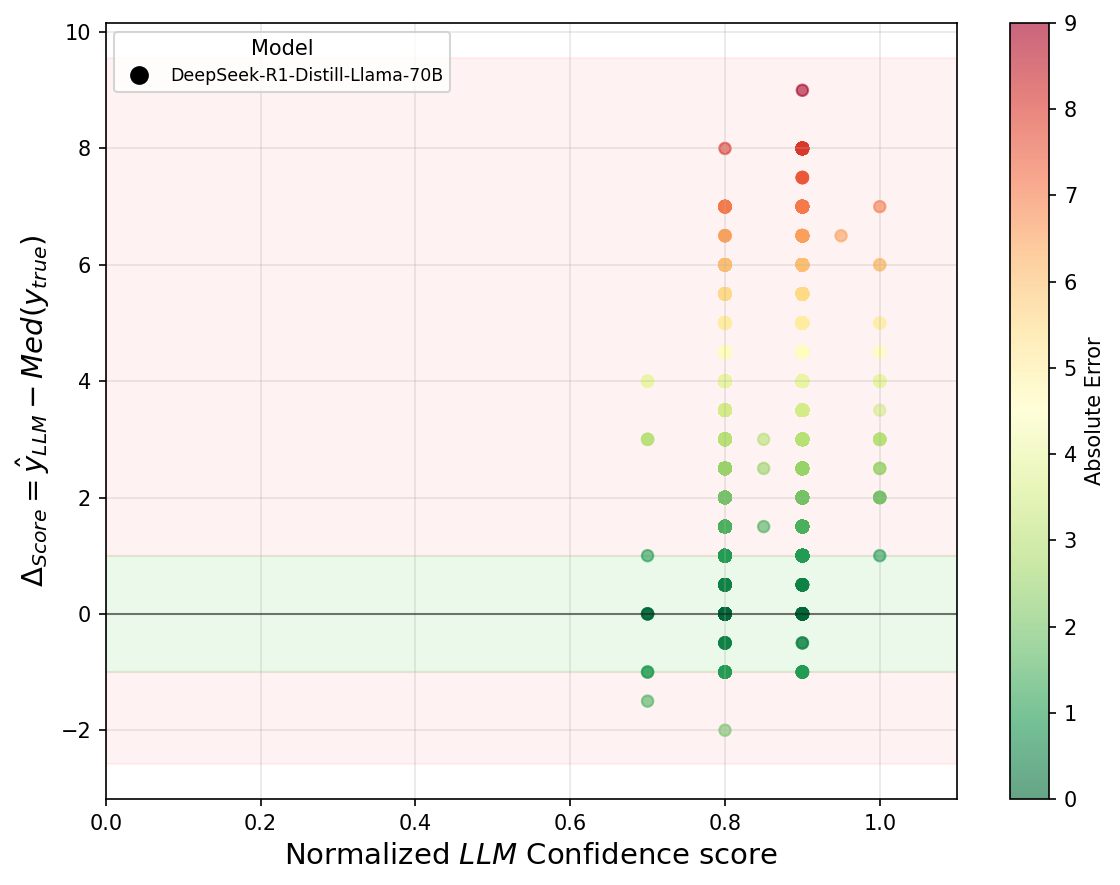} }}%
    \qquad
    \subfloat[\centering gpt-oss-120b]{{\includegraphics[width=0.45\linewidth]{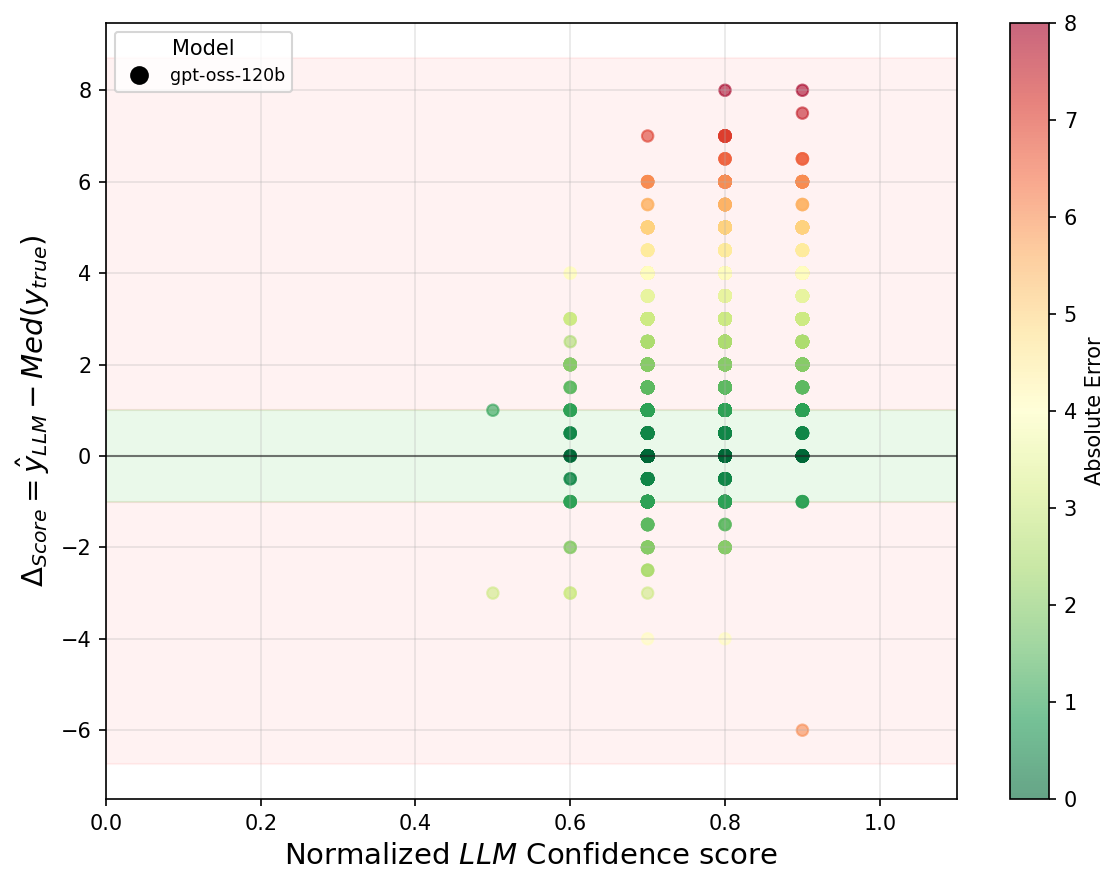} }}%

    \caption{LLM-reported Confidence vs Prediction Bias when reviewing complete content presented in a scientific article using the full-text}
    \label{fig:LMRSD_exp2_conf_plot}
\end{figure}

Fig \ref{fig:LMRSD_exp1_conf_plot} presents the confidence-error plot for the idea review task, where models assess paper ideas based solely on titles, abstracts, and keywords. When operating on this limited information, models display broader confidence distributions spanning 0.5 - 1.0, suggesting minor awareness of input constraints. Qwen3-32B and Llama-3.3-70B-Instruct produce predictions with confidence below 0.6 in non-trivial proportions, while Gemma-3-27B-IT (Fig \ref{fig:LMRSD_exp1_conf_plot_remaining}) exhibits tighter concentration around 0.7 - 0.9. Despite this spread, the fundamental pattern here is a green zone sparsely populated even at moderate confidence levels (0.6- 0.8), and the majority of predictions cluster in the red region above $\Delta_{\text{Score}} = +1$, indicating systematic overestimation by 2 - 8 points. \textbf{Qwen3-32B}, achieved the lowest MAE and RMSE for idea evaluation (Table \ref{tab:exp1_metrics}), shows a distributed confidence profile with some negative values (underestimations), though overestimation remains dominant. \textbf{Gemma-3-27B-IT} displays overconfidence while \textbf{Llama-3.3-70B-Instruct} appears most conservative, with $\Delta_{\text{Score}}$ rarely exceeding and somewhat cleaner separation between error magnitudes across confidence levels. Across all models evaluated in Experiment 1 (Fig \ref{fig:LMRSD_exp1_conf_plot_remaining}), the confidence-error plots convey a consistent message about the limited value of assessing ideas based solely on abstracts, and while models express more varied confidence than in full-text conditions, this variation does not meaningfully predict accuracy.

Reasonably, providing full text should lead to improvements in both accuracy and calibration, since LLMs are given the methodology, results, and other sections, rather than just the abstract. Paradoxically, Fig \ref{fig:LMRSD_exp2_conf_plot} shows full-text access \textit{increases} verbalized confidence, but substantial errors remain. Comparing the two experimental conditions, we can notice a clear shift emerge. In abstract-only reviews, confidence distributions showed a meaningful spread (0.5 - 1.0) with some low-confidence predictions. With full-text access, confidence collapses into a narrow band at 0.7 - 1.0 across all models, with very few predictions below 0.6. This simultaneously extends further into positive territory, producing extremely positive outliers for some models concentrated in 0.8 - 0.9 confidence range. The inclusion of reasoning models (DeepSeek-R1-Distill-Llama-70B, Qwen3-Next-80B-A3B-Thinking, gpt-oss-20B, gpt-oss-120B) in Experiment 2 provides an opportunity to test whether intermediate tokens (chain-of-thought reasoning traces) improve calibration. As shown in Table \ref{tab:exp2_metrics}, \textbf{gpt-oss-120B} achieved the lowest MAE among reasoning models, yet its confidence-error plot reveals substantial overestimation with confidence highly concentrated at 0.8 - 0.95. The reasoning models show no stark improvement in the joint distribution structure compared to dense models. Using "thinking" models post-trained for reasoning, safety, and performance does not translate to better calibrated uncertainty expression. The full set of confidence-error plots for Experiment 2 (Fig \ref{fig:LMRSD_exp2_conf_plot_remaining}) reinforces this pattern across all nine models with concentration of confidence values at 0.7 - 1.0 regardless of where the predictions fall on the plot. Models appear to leverage full-text information primarily to express greater certainty rather than greater accuracy.

If we return to our motivating comparison with human reviewers (Fig \ref{fig:LMRSD_data_review}), this visual analysis revealed a fundamental structural difference. Human reviewers produce a joint distribution plot with meaningful covariation in confidence across the range $1$ to $8$, reflecting context-dependent epistemic states. LLMs as reviewers collapse this structure into a narrow vertical band with high verbalized confidence (0.7--1.0) stretched across $\Delta_{\text{Score}}$ values ranging from accurate ($\pm 1$) to grossly overestimated ($+8$ to $+10$). Whether this reflects a failure of instruction tuning, an inherent limitation of LLM verbalized uncertainty, or simply that the models lacking this capability remains an open question. What is clear is that current LLMs, including frontier reasoning models with explicit chain-of-thought capabilities, cannot reliably distinguish between judgments they should trust and those they should doubt. For a system operating in high-stakes scientific gatekeeping, this absence of calibrated uncertainty expression may be as consequential as the accuracy limitations documented in previous sections.

\subsection{Impact of LLM Reviews on Publication Outcomes}
\label{RQ3}
\begin{figure}[!htb]
    \centering
    {\includegraphics[width=0.85\linewidth]{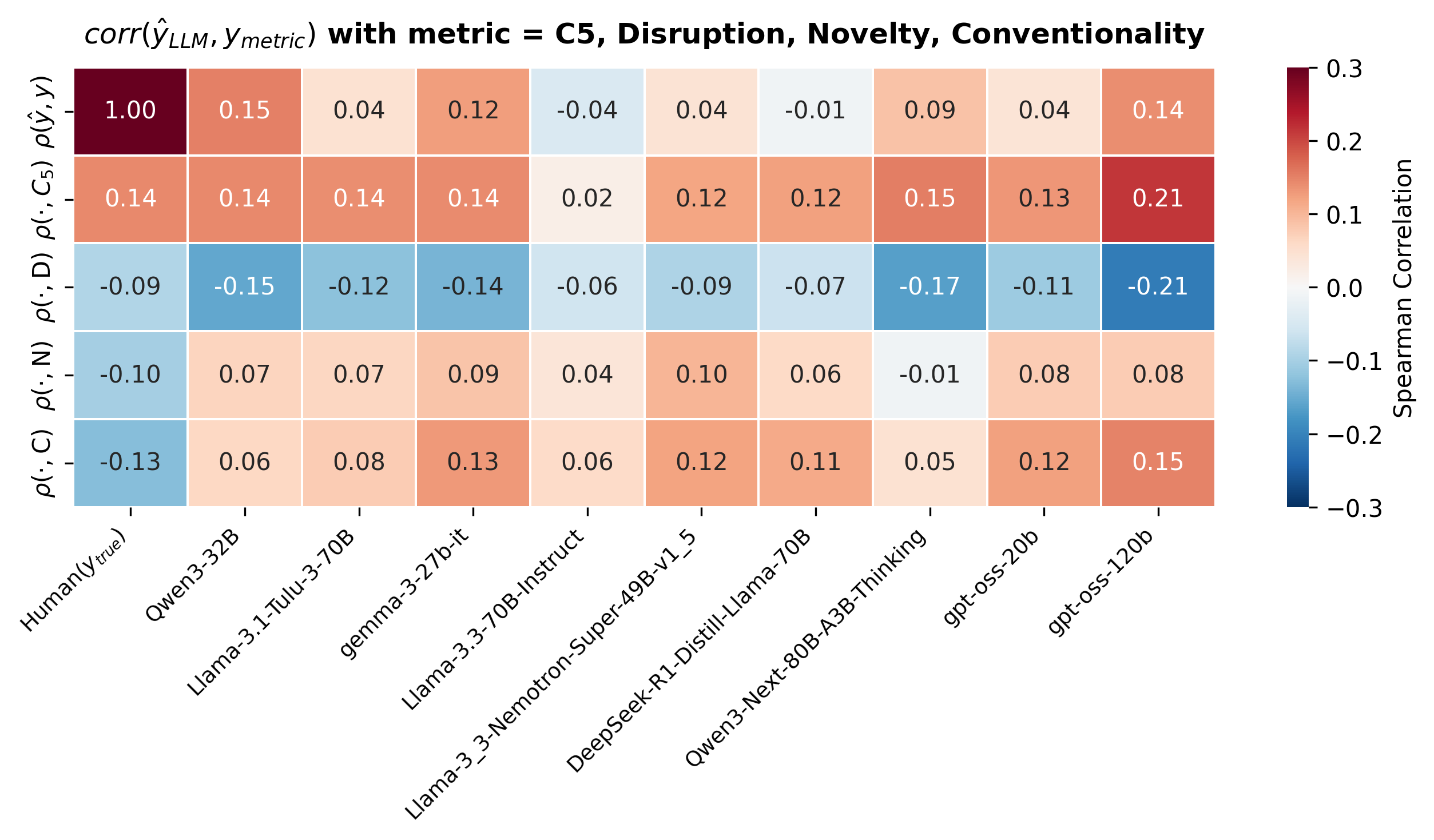} }
    \caption{Correlation for post publication impact metrics against all models $\hat{y}_{LLM}$}
    \label{fig:LMRSD_exp4_thresholds}%
\end{figure}

Capturing the impact of LLM generated reviews is not a trivial task because beyond the content and score of the review there is no guide to search for meaningful signals. These signals ideally should validate the usefulness or utility of LLM generated reviews on the article outcomes and research community in general. To aid us in this analysis, we prepared a dataset suitable for the task $D_{LMRSD.c}$ which includes scientific articles and other meta information about reviews observed in $D_{LMRSD.a}$ $\&$ $D_{LMRSD.b}$ along with science of science metrics namely: a.) \textit{hit pct} status, a binary variable capturing if a paper received the top $n$ percentage of citations in the respective field, so $\text{hit}-1 \text{pct}, \text{hit}-5 \text{pct}, \text{hit}-10 \text{pct}$; b.) \textit{novelty} \& \textit{conventionality}, which is a pre-calculated metric telling us the Median and 10 pct $Z$ scores assessing at publication time the atypicality and conventionality of ideas/knowledge combinations; and (3) \textit{disruptiveness}, which also is a pre-calculated metric telling us on a range of $[-1, 1]$ whether a given paper is disruptive $(D=1)$ or developing $(D=-1)$. Figure \ref{fig:LMRSD_exp4_thresholds} shows the correlation heatmap comparing ground truth reviewer scores against LLM generated scores. To preface our earlier findings on weak LLM-human correlations, we included them in the first row ($\rho(\hat{y}, y)$) and broadly correlation across all models for the publication metrics.

For 5-year citation counts ($C_5$), several LLMs demonstrate correlations \textit{comparable to or exceeding} those of ground-truth scores. While this should not be interpreted as predicting eventual scholarly impact, it can be viewed as a medium to long-term signal of how extensively a scholarly paper is discussed within the citation network. Ground-truth human scores correlate with $C_5$ at $\rho = 0.14$, whereas gpt-oss-120b achieves $\rho = 0.21$, representing a 50\% increase over human scores. Correlations about the disruptive index show a consistent negative relationship across all values, indicating that both humans and LLMs tend to assign lower scores to more disruptive work. This pattern is more prevalent in LLMs, with gpt-oss-120b showing the strongest negative correlation ($\rho = -0.21$) compared to human reviewers ($\rho = -0.09$). Disruptive index tells us on a scale of $-1$ to $1$, whether a given scholarly work is cited mimicking the behavior of disruptive work or developing/consolidating work. One potential interpretation here is LLMs and humans alike are more than likely to identify consolidating work and its rather hard to identify disruptive work. The novelty correlations show weak correlation values for most LLMs ($\rho \approx 0.04-0.10$), whereas human scores exhibit a weak negative correlation ($\rho = -0.10$). Similarly, for conventionality, LLMs show positive correlations ($\rho \approx 0.05-0.15$) while human reviewers show a negative relationship ($\rho = -0.13$). Overall, if we were to use the $\hat{y_{LLM}}$ or $y_{true}$ correlations with the science metrics outlined above we can definitely say there are contrasting patterns and they suggest that LLMs and humans may evaluate scholarly contributions through fundamentally different lenses with LLMs potentially favoring work that combines familiar elements in detectable patterns.

\begin{figure}[!htb]
    \centering
    \subfloat[\centering Hit paper AUC (LLM-vs-Human)]{{\includegraphics[width=0.45\textwidth]{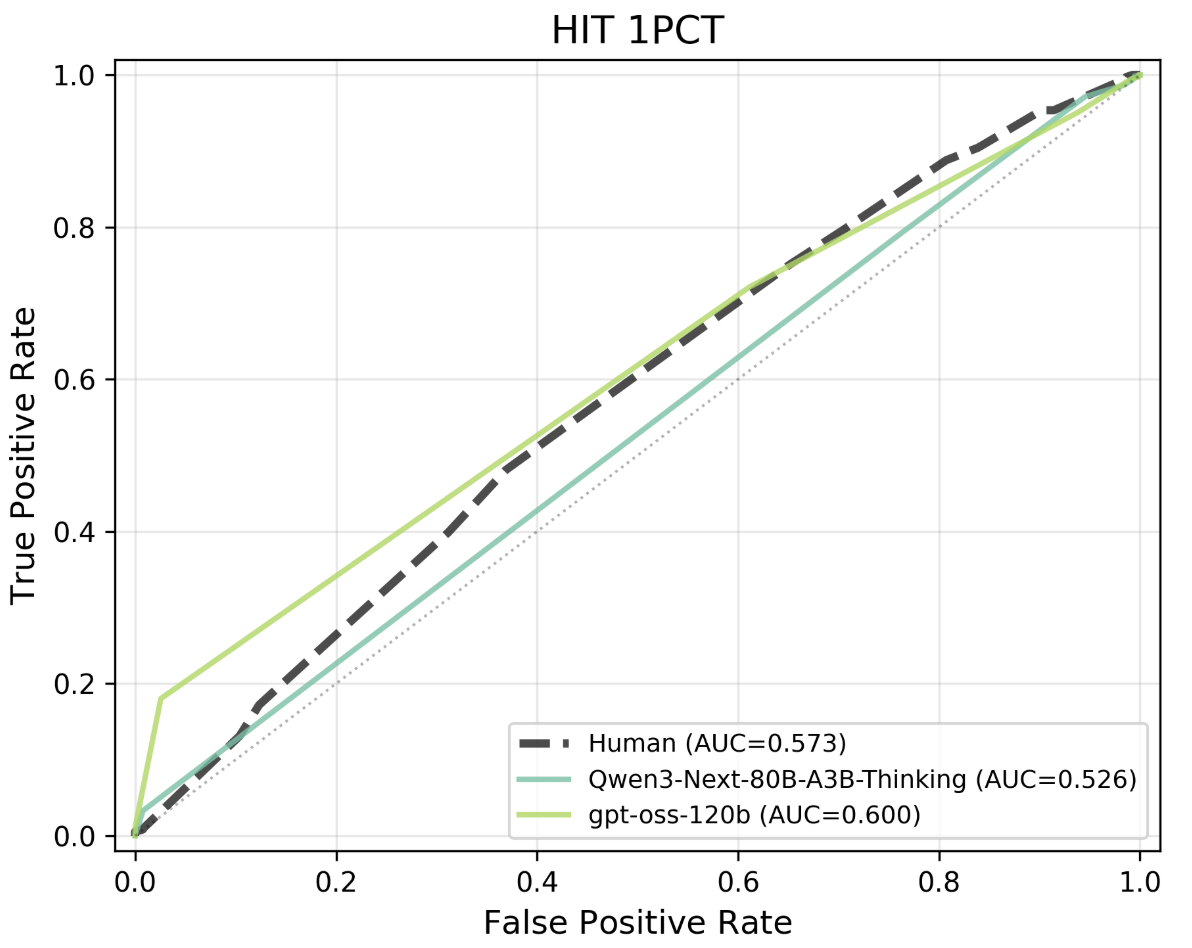} }}%
    \qquad
    \subfloat[\centering Hit 1 \% paper-review score thresholds for reasoning models]{{\includegraphics[width=0.45\textwidth]{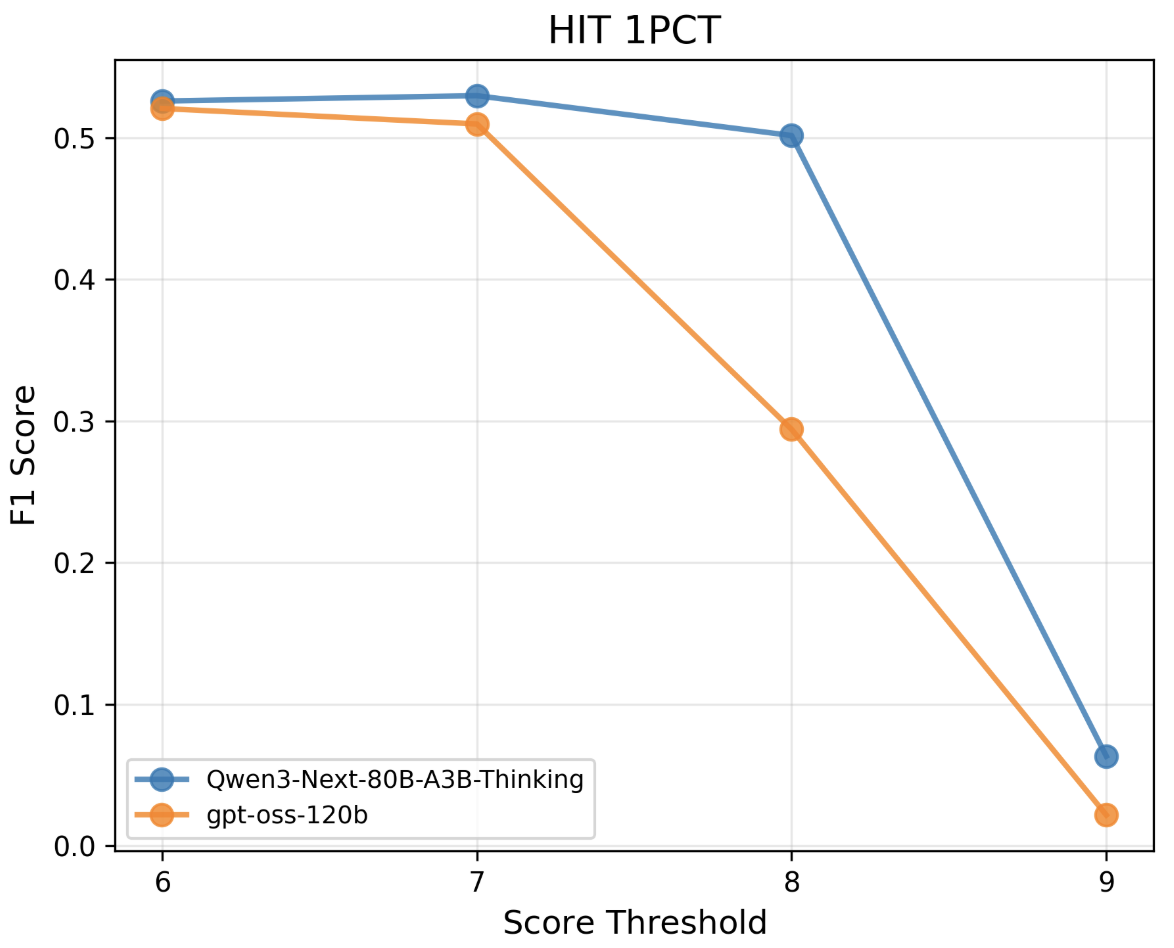} }}%
    \qquad
    \subfloat[\centering Hit 5 \% paper-review score thresholds for reasoning models]{{\includegraphics[width=0.45\textwidth]{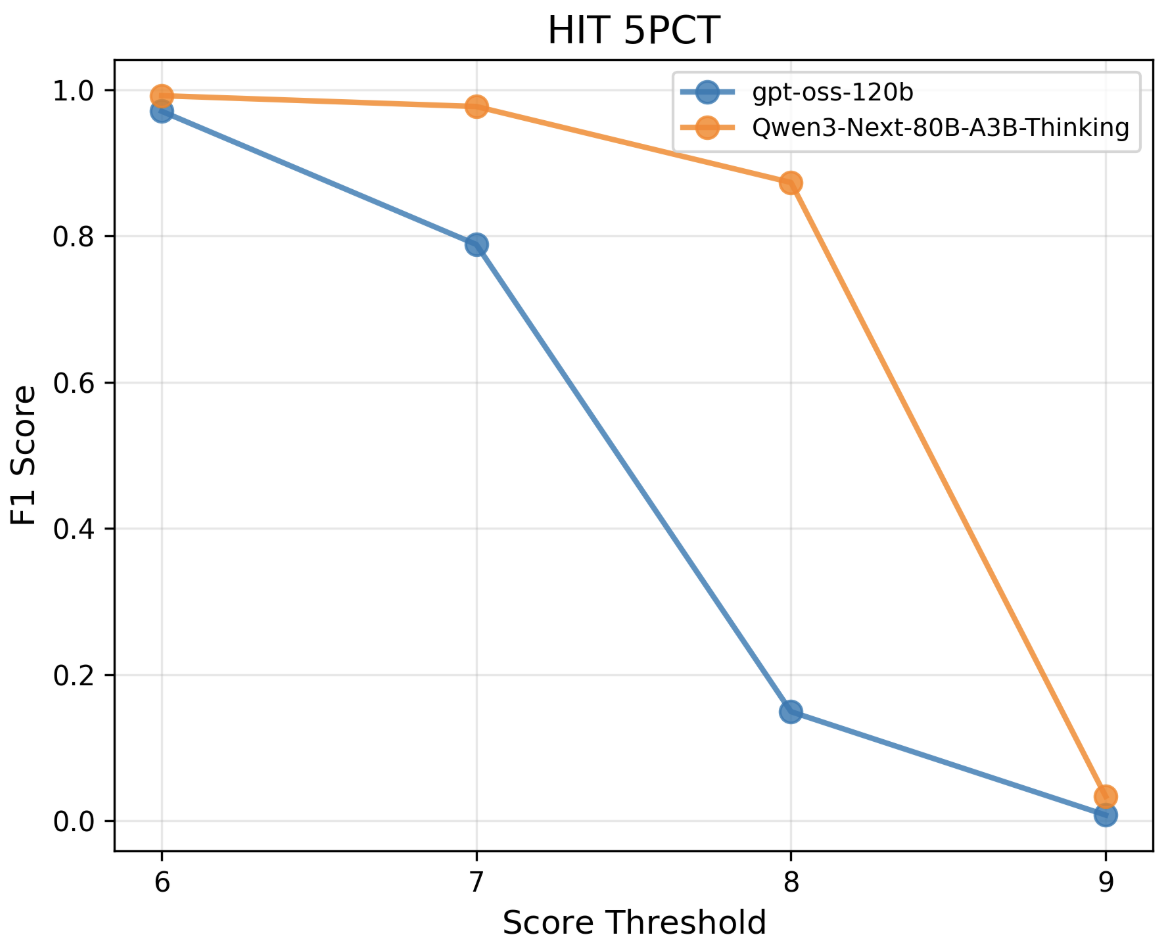} }}%
    \qquad
    \subfloat[\centering Hit 10 \% paper-review score thresholds for reasoning models]{{\includegraphics[width=0.45\textwidth]{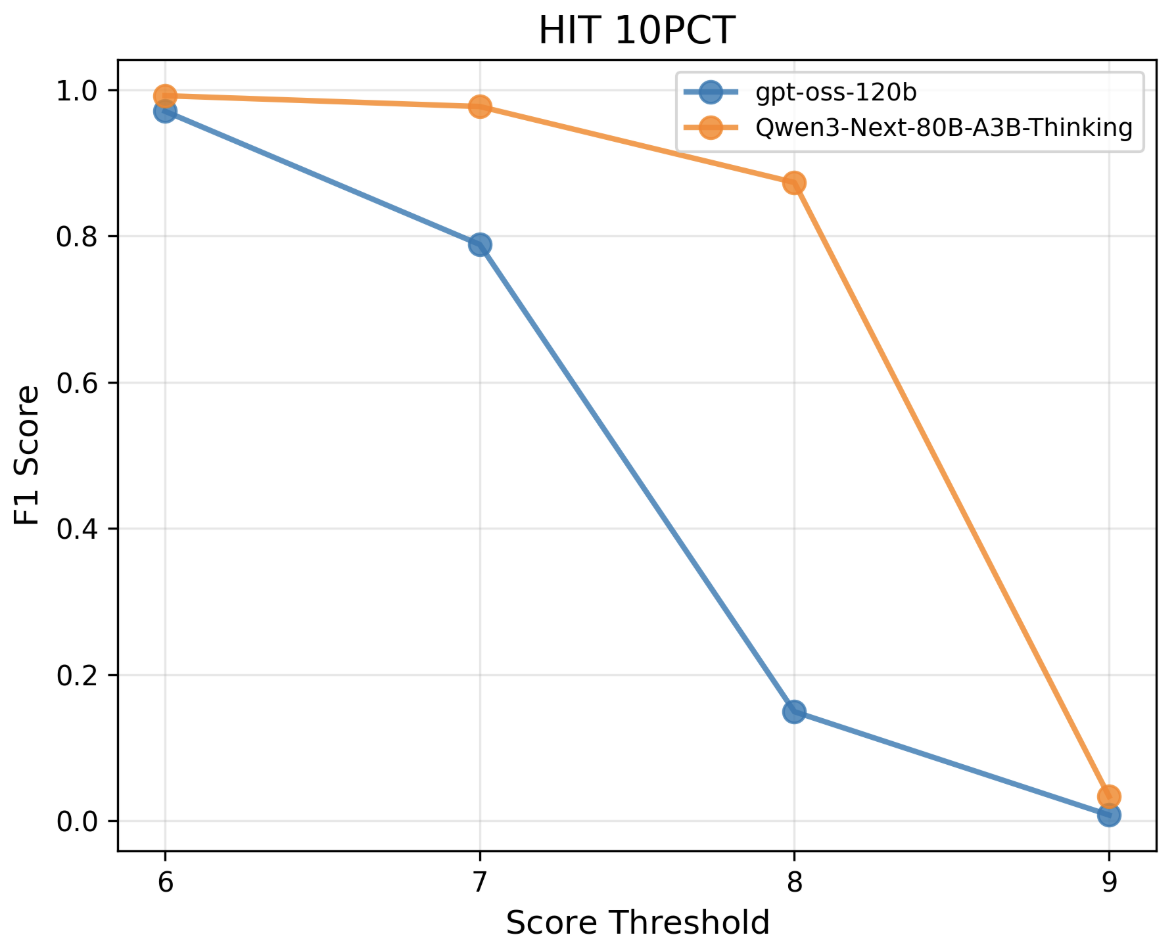} }}%

    \caption{Comparing the hit paper predictive thresholds for LLMs on post-publication outcomes dataset $D_{LMRSD.c}$ }
    \label{fig:LMRSD_exp4_hit_outcomes}
\end{figure}

Another mechanism to assess the usefulness of LLM reviews is identifying how well LLM reviews identify high-impact research. So, we evaluated their ability to predict hit papers using the dataset $D_{LMRSD.c}$, which includes papers stratified by three categories: true positives (hit papers correctly identified by humans), false negatives (hit papers missed by human reviewers), and false positives (non-hit papers rated highly by humans). We are aware of this adversarial nature of classification since \textit{true/false} or \textit{positive/negative} is entirely being constructed based on the empirical observation. Figure \ref{fig:LMRSD_exp4_hit_outcomes}(a) shows ROC curves comparing human reviewers against top two reasoning models for hit-1\% prediction. The gpt-oss-120b model achieves an AUC of 0.600, marginally outperforming human reviewers (AUC = 0.573), while Qwen3-Next-80B-A3B-Thinking underperforms at AUC = 0.526. At face value, this tells us a model with capabilities like gpt-oss-120b can match or slightly exceed human expert ability in identifying papers that will achieve top-1\% citation status in their field.  Figures \ref{fig:LMRSD_exp4_hit_outcomes}(b-d) different score thresholds examining F1 scores for hit paper classification. For hit-1\% papers, both models achieve comparable F1 scores ($\sim0.52-0.53$) at lower thresholds ($\leq 7$), but performance varies substantially at higher thresholds. At threshold 8, gpt-oss-120b drops to F1 $\approx$ 0.30 while Qwen3-Next model maintains F1 $\approx$ 0.50. For hit-5\% and hit-10\% identification, the pattern reverses: Qwen3-Next demonstrates remarkable robustness, maintaining F1 $>$ 0.87 at threshold 8, while gpt-oss-120b experiences precipitous degradation (F1 $\approx$ 0.15).  This asymmetry in F1 score thresholds is helpful to weigh the impact of scores generated by models in specific regions of the score distribution since they may carry a useful predictive signal.

Taking a step back and looking at all of these findings, its easy to observe a paradoxical pattern when evaluating usefulness LLMs as peer reviewers especially on post-publication outcomes. One way to look at these findings is thinking about the peer review process, whether conducted by humans or models, as a systematic long-horizon assessment asking for several demanding qualities. Despite showing weak correlation with human reviewers ($\rho < 0.16$), LLMs show stronger associations with certain post-publication outcomes than human reviews themselves. One interpretation is that LLMs and humans evaluate orthogonal dimensions of scientific quality: humans assess methodological rigor, novelty of claims, and contextual significance, while LLMs may detect textual patterns correlated with eventual citation success.  The in-distribution training tasks and capabilities allow LLMs to easily gravitate to clarity of exposition, breadth of literature used, or alignment with trending research topics.  Alternatively, this disconnect may reflect known biases in citation based metrics themselves, which reward incremental work in active subfields over disruptive or novel contributions, because they take longer to be recognized.

\section{Ablations}
Our main experiments showed overconfidence, inflation, and weak alignment across all evaluated models. Naturally, these findings raise questions about the underlying mechanisms driving these results. In this section, we present results from three targeted ablations that understand the above behaviors. First, we examine generation perplexity to determine whether models are genuinely uncertain when producing reviews or whether they confidently generate miscalibrated outputs. Next, we analyze the similarity between human-written reviews and LLM-generated reviews to understand the broader semantic representations of the review texts. Lastly, we investigate whether increased reasoning effort with prompt engineering can mitigate the observed biases in review scores. Due to the ablations experimental structure, we only used gpt-oss-120b, and Qwen3-Next-80B-A3B-Thinking models on the $D_{LMRSD.c}$ dataset.

\subsection{Ablation 1: How confident are the frontier of models when it comes to predicting peer-review scores and generating peer review descriptions ?}

Generating reviews for scientific texts is not a stylistically different task in terms of linguistic fluency or task familiarity for language models. While our results jointly observed in Table \ref{tab:exp2_metrics} and Fig. \ref{fig:LMRSD_exp2_conf_plot} suggested that the frontier of language models are over-confident when it comes to plotting the self-reported confidence scores against the comparison of difference between ground truth and LLM generated scores, it was still peculiar behavior because the task of peer-review was in the domain of model's capabilities.

\begin{figure}[!htb]
    \centering
    {\includegraphics[width=0.75\linewidth]{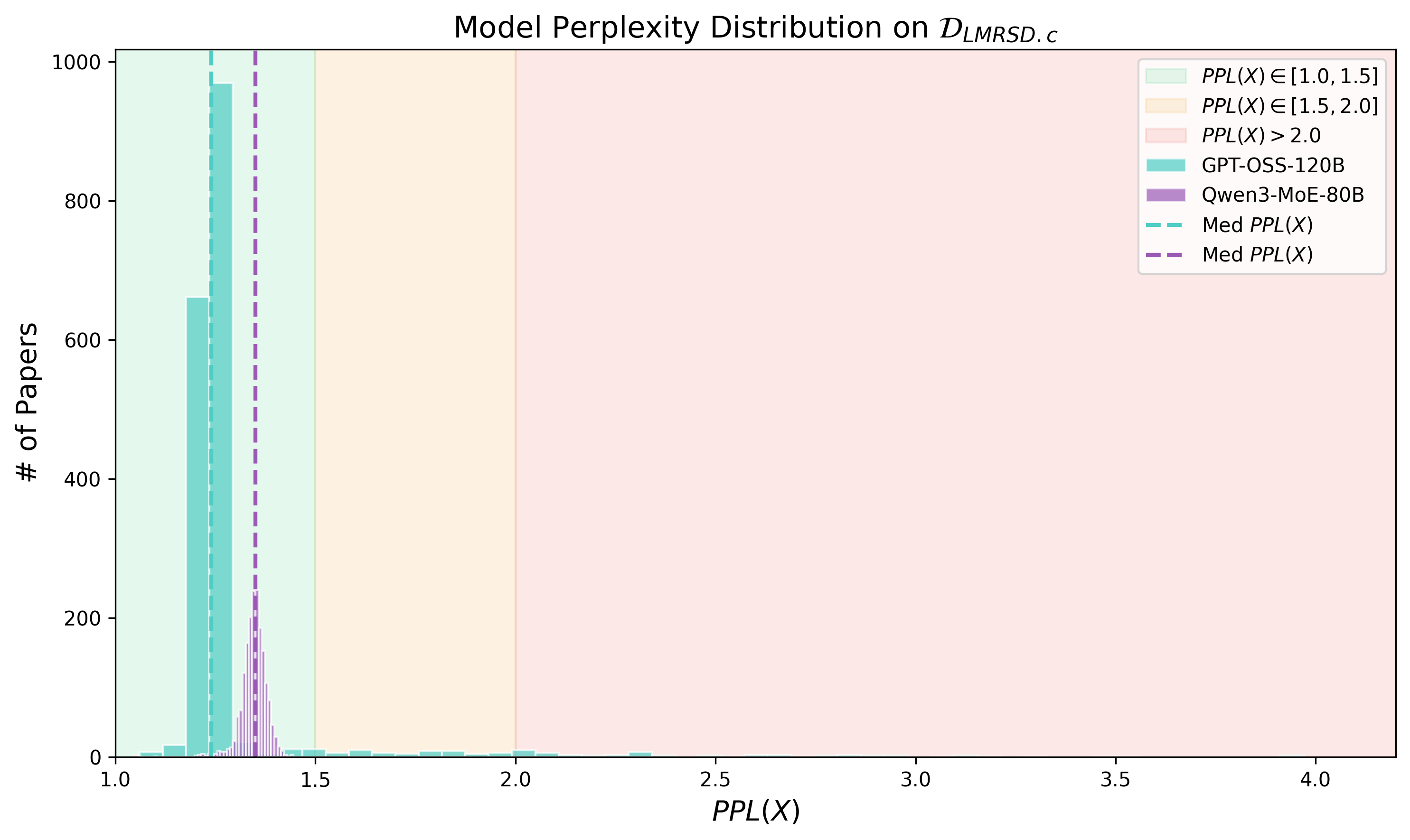} }
    \caption{Distribution of Perplexity Scores for best performing models on post-publication outcomes dataset $\mathcal{D}_{LMRSD.c}$}
    \label{fig:lmrsd_abl1_pplx}%
\end{figure}

To explore more about the model's overconfidence in peer review, we analyzed the perplexity scores of the two top-performing reasoning models \textbf{gpt-oss-120b}, and \textbf{Qwen3-Next-80B-A3B-Thinking}. Figure~\ref{fig:lmrsd_abl1_pplx} and Table~\ref{tab:perplexity} present generation perplexity for two reasoning models on $D_{LMRSD.c}$. Both models exhibit remarkably low perplexity concentrated in the $PPL(x) \in [1.0, 1.5]$ zone (gpt-oss-120B: median 1.24; Qwen3-Next-80B-A3B-Thinking: median 1.35) for nearly 95\% of the samples, indicating that peer review generation is nearly deterministic on average, fewer than 1.4 plausible next tokens are considered at each step. While Qwen3-Next-80B-A3B-Thinking  maintains an exceptionally tight distribution entirely within the $[1.0, 1.5]$ zone, GPT-OSS-120B exhibits a longer tail extending into moderate and high uncertainty regions, with 6.2 \% of papers exceeding $PPL(X) \geq 1.6$. This suggests gpt-oss-120B occasionally encounters content that deviates from familiar review patterns, while Qwen3-Next-80B-A3B-Thinking treats the task as uniformly predictable. These low perplexity scores reveal that these models have acquired strong priors over the form of peer review—the linguistic patterns, structural conventions, and stylistic norms. However, this confidence and fluency in review generation does not translate to accuracy in review scoring. Specifically, models confidently produce well-formed reviews while systematically miscalibrating their numerical scores.

\begin{table}[t]
\centering
\caption{Model Familiarity with Peer Review Generation task | Ablation-1}
\label{tab:perplexity}
\begin{tabular}{lcc}
\toprule
\textbf{Metric} & \textbf{GPT-OSS-120B} & \textbf{Qwen3-MoE-80B} \\
\midrule
Median $PPL(X)$ & 1.24 & 1.35 \\
IQR & [1.23, 1.26] & [1.33, 1.37] \\
Std. Dev. & 0.26 & 0.03 \\
P5 / P95 & 1.21 / 1.71 & 1.30 / 1.39 \\
\midrule
Mean LogProb/Token & -0.22 & -0.29 \\
Avg. Output Length & 3,215 tokens & 1,288 tokens \\
\bottomrule
\end{tabular}
\vspace{1mm}
\end{table}

\subsection{Ablation-2: How similar/dissimilar are the reviews written by top language models compared to reviews generated by human reviewers?}

Beyond confidence in generating reviews, we wanted to understand whether the review texts generated by the language models are semantically similar to those generated by humans. We used ground-truth reviews and language model-generated reviews to generate embeddings with \texttt{Qwen3-Embedding-0.6B} and ran them through a softmax to capture probability distributions. We quantified similarity and divergence between human-generated and LLM-generated reviews using cosine similarity, Kullback-Leibler (KL) divergence in both directions, and Jensen-Shannon (JS) divergence.  

Table \ref{tab:abl1_review_similarity} clearly shows both models achieving high cosine similarity with human reviews, with mean values close to 0.8, indicating that LLM-generated reviews are largely semantically similar with human-generated reviews when we use cosine similarity as a basis. Since KL Divergence is an asymmetric metric, we capture $D_{KL} (y_{true}, \hat{y}_{LLM})$ and $D_{KL} (\hat{y}_{LLM}, y_{true})$ to capture the divergences shown by LLM generated reviews against human content. gpt-oss-120b shows a lower bi-directional divergence value suggesting it maps human patterns well and avoids generating information unlikely under the human ($y_{true}$) distribution. The overall distribution similarity captured from JS score tells us that gpt-oss-120b reviews are not just semantically similar, but from an information theory perspective they show a higher distribution fidelity to human content and the same cannot be said for Qwen3-Next-80B-A3B-Thinking.

\begin{table}[t]
\centering
\caption{Embedding-space similarity between LLM generated reviews and ground truth human peer reviews on $D_{LMRSD.c}$ dataset. | Ablation 2}
\label{tab:abl1_review_similarity}
\begin{tabular}{lcccc}
\toprule
Model & Cosine & KL($y_{true}$ $\to$ $\hat{y_{LLM}}$) & KL($\hat{y_{LLM}}$ $\to$ $y_{true}$) & JS \\
\midrule
gpt-oss-120b    & \textbf{0.8770 $\pm$ 0.0912} & \textbf{0.8210}          & \textbf{0.9114}          & \textbf{0.1448} \\
Qwen3-Next-80B-A3B-Thinking    & 0.8764 $\pm$ 0.0270          & 1.1635                   & 2.1133                   & 0.2478 \\
\bottomrule
\end{tabular}
\vspace{1mm}
\end{table}

\subsection{Ablation 3: Does reasoning strength combined with adjusted prompt considerations deviate the peer-review scores of reasoning models ?}

\begin{figure}[!htb]
    \centering
    {\includegraphics[width=0.95\linewidth]{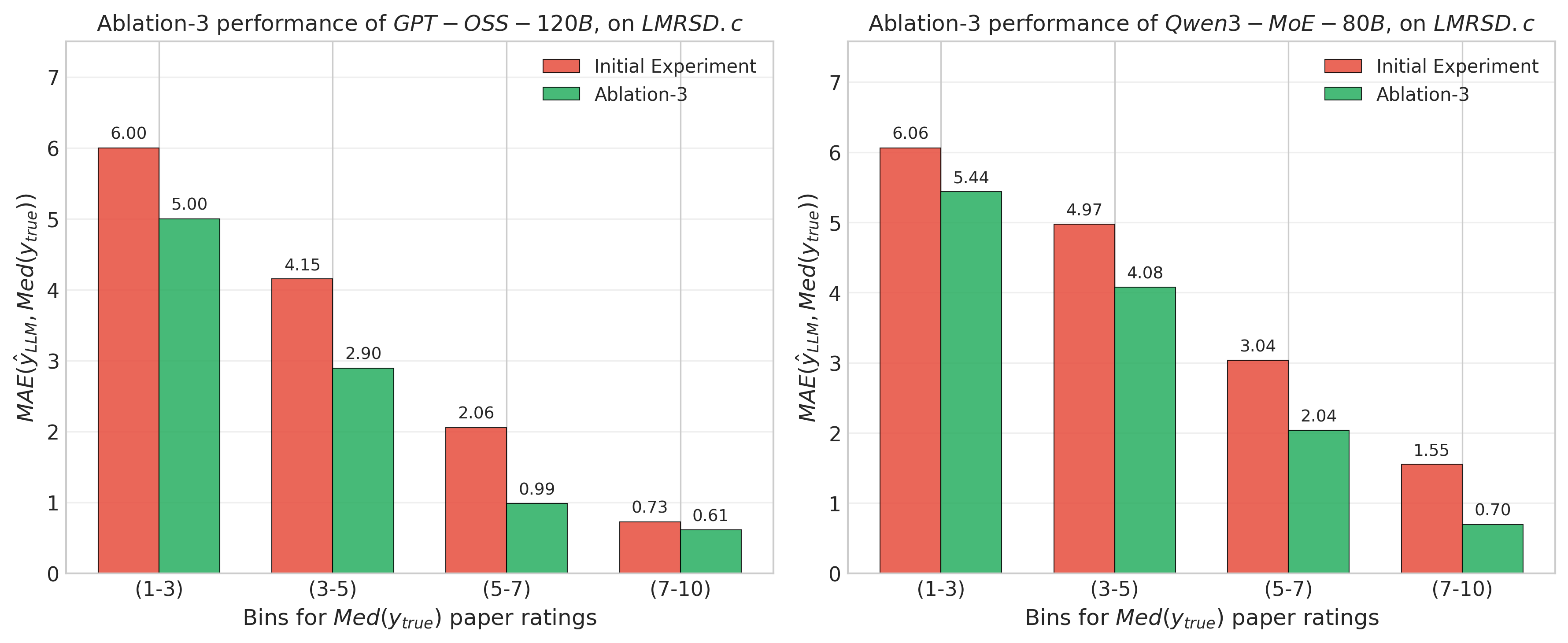} }
    \caption{MAE stratified by ground truth quality bins before and after applying the scoring calibration prompt on post-publication outcomes dataset $\mathcal{D}_{LMRSD.c}$}
    \label{fig:lmrsd_abl3_mae}%
\end{figure}

Our prompts play a central role in enabling models to produce peer review scores for the vast majority of papers. Naturally, this raises the question of the impact one would tie to the inclusion of nuanced considerations in the input prompt. Additionally, most of the initial experiment results placed no emphasis on \textit{reasoning effort} or set a reasoning budget, but rather just set an overall $8k$ output max token limits to efficiently process the entire $D_{LMRSD.b}$ dataset. So, we've changed the reasoning effort to \textbf{high}, set the maximum output tokens to $36k$, and updated the input prompt as observed in Appendix \ref{zero_shot_ablations_prompts} to understand if explicit scoring calibration instructions combined with increased reasoning effort will reduce inflation of scores across bins. We ran the inference on $D_{LMRSD.c}$ and capturing the before and after for \textbf{gpt-oss-120b} and \textbf{Qwen3-Next-80B-A3B-Thinking} helped us notice that there is a statistically significant reduction in MAE, and RMSE for both the models. The updated prompt (Appendix \ref{zero_shot_ablations_prompts}) and high reasoning effort computed on matched paper pairs show an overall change from baseline. The ablation results in Table. \ref{tab:ablation3} shows an overall negative delta indicating improvement for MAE, RMSE values for both the models. 

\begin{table}[t]
\centering
\caption{Impact of scoring prompt and reasoning effort on post-publication outcomes dataset $D_{LMRSD.c}$ | Ablation 3}
\label{tab:ablation3}
\begin{tabular}{llcccc}
\toprule
\textbf{Model} & \textbf{Condition} & \textbf{N} & \textbf{MAE} & \textbf{RMSE}\\
\midrule
\multirow{3}{*}{gpt-oss-120B} 
 & Initial Experiment & 339 & 1.882 & 2.344 \\
 & Ablation 3 & 339 & 1.130 & 1.549 \\
 & $\Delta$ & --- & \textbf{-0.752} & \textbf{-0.795} \\
\midrule
\multirow{3}{*}{Qwen3-Next-80B-A3B-Thinking} 
 & Initial Experiment & 976 & 2.784 & 3.086 \\
 & Ablation 3 & 976 & 1.849 & 2.234 \\
 & $\Delta$ & --- & \textbf{-0.935} & \textbf{-0.852} \\
\bottomrule
\end{tabular}%
\vspace{1mm}
\end{table}

Fig. \ref{fig:lmrsd_abl3_mae} shows difference in absolute error across bins before and after the ablation. Clearly, the effect of the ablation is felt in the lower score bins $1-3$ and $3-5$ as they observe the largest absolute improvement where baseline models exhibited the most severe overestimation. While it is difficult to attribute the causal effect of the improvement in MAE scores, the clear takeaway here is the practical suggestion of including explicit calibration instructions to mitigate LLMs when reviewing potentially weaker submissions. While we didn't record specific metrics to study the effect of reasoning effort, the performance of \textbf{gpt-oss-120b} and \textbf{Qwen3-Next-80B-A3B-Thinking} on GPQA with \textbf{high} reasoning acted as a motivating factor to allow these models extend their thinking traces before generating peer-review descriptions and scores. The results from this ablation only show the benefit of explicit calibration instructions over naive prompting when it comes to LLM generated reviews. While it doesn't improve alignment with human judgments, the persistent compression of the score distribution suggests that prompt engineering alone is insufficient to achieve human-level calibration, motivating future work on fine-tuning or customized RL environments that utilize a broader set of tools to generate peer reviews.

\section{Discussion}

Our experiments focused on understanding the utility and broader safety risks involved in fully/completely automating peer reviews with language models. Anecdotally, it is common knowledge that most researchers run \textit{pre-reviews} of manuscripts through commercial providers such as ChatGPT\footnote{https://chatgpt.com}, Claude\footnote{https://claude.ai}, Gemini\footnote{https://gemini.google.com} and other popular platforms for a wide range of peer review related tasks. Our findings suggest that current frontier LLMs, despite their capabilities in other domains, are not suitable for autonomous peer review in their present form. Their systematic grade inflation and inability to identify low-quality work would compromise the normal functioning of peer review. While our analysis is focused on open-weight frontier models, a glimpse of the pitfalls can be abstracted to every model at least from a model safety standpoint. Perhaps the most surprising finding of our study is the limited gap between dense and reasoning models acting as peer-reviewers. While it contradicts current benchmarks, it's worth noting that our experiments are subtle, subjective and chain-of-thought based post-training datasets heavily cater to verifiable domains such as math/coding.

Our attempt to clearly outline the potential pitfalls and supplement our experimental analysis with \ref{RQ3} is rooted in the belief that there are some potential benefits of including language models in the broader pipeline of automating scientific reviews. Specifically, the performance of gpt-oss-120b on $D_{LMRSD.c}$ tells us that humans and LLMs are potentially looking at orthogonal dimensions when assessing science. While its difficult to speculate about human priors, we can certainly say LLMs are checking surface-level textual patterns, clarity of written works, literature engagement, or in some cases match with trending research topics.  This interpretation gains support from our analysis of disruption and novelty metrics since humans and LLMs alike show negative correlations with disruptive index, but LLMs exhibit this bias more strongly ($\rho = -0.21$ for gpt-oss-120b versus $\rho = -0.09$ for humans). From a science of science perspective, this finding echoes concerns about the Matthew Effect \citep{merton1968matthew} in science \citep{algaba2025large} and the documented tendency of peer review to favor conventional over innovative research \citep{lee2013bias, boudreau2016looking}. If LLMs amplify these conservative biases, their deployment could accelerate the marginalization of disruptive ideas precisely when scientific progress depends on paradigm shifts.

Current alignment methods rely on human preference data to train models, implicitly assuming that human judgments provide optimal training signals. Through our experiments and ablations, we argue for broader evaluations of LLMs specifically mimicking human cognitive processes when conducting peer-review to maximize the utility of LLMs in scientific evaluation process. It is evident from our findings that peer-review is a long-horizon agentic task that requires specialized tools, vast literature databases for accurate peer reviews. \textit{PaperBench} \citep{starace2025paperbenchevaluatingaisability} is a great benchmark that is a step in the right direction of creating specialized agentic RL environments where LLMs attempt to replicate AI research. A broader dataset across computational science will yield agentic reviewers optimized for identifying not just citable work but genuinely transformative contributions.

Our findings suggest that pure text-based review fundamentally limits LLM capabilities. Human reviewers routinely consult literature databases, run provided code, examine figures in detail, and draw on community knowledge unavailable in the manuscript. Integrating these tools into agentic review systems, where LLMs can query databases, execute code, and reason over structured scientific knowledge, could address limitations inherent to text-only approaches. Peer review as a task may need to be decomposed into sub-tasks with different alignment requirements. Following insights from \citep{staudinger-etal-2024-analysis} on the multi-dimensional nature of peer review, these benchmarks must include rubrics to check methodological soundness, novelty assessment, clarity evaluation, and reproducibility checking, each including a diverse distribution of samples rather than highly influential samples. Developing benchmarks that include appropriate harsh feedback in scientific evaluation contexts, while preserving constructive critiques, represents a novel alignment challenge. Having a diversity of papers (quality wise) could ensure LLM reviews limit sycophancy \citep{perez2022discovering, cheng2025social}, and project a willingness to assign lower scores. 

\section{Conclusion \& Future Work}
Frontier LLMs are highly capable at generating fluent, structurally appropriate peer reviews (evidenced by low perplexity and high semantic similarity to human reviews), yet remain systematically different in their review scores. The entire process of a peer-review is a long-horizon task where the cognitive abilities of the human are put to the test, and during this multi-hop process contextual understanding of the scientific work along with internal priors about the subject dictate the effectiveness of the review. Thematically, this peer-review process maps well as a task for most frontier LLMs. To be more specific, the peer-review agent must be multi-modal, and equipped with tool use (internet search, run code) to capture different aspects of the peer-review. A demonstration of this future-direction can be observed in the Appendix \ref{agentic_peer_review} which outlines the planning, task composition, and summarizing the results.

\section{Limitations}
We acknowledge seemingly obvious limitations of our study and point them towards some open questions. Firstly, we do note that our analysis primarily focuses on computational science venues such as ICLR, NeurIPS, CoRL, MIDL. Whether these findings generalize to other disciplines, particularly those with different methodologies or experimental structure, remains unknown. Our human baseline relies on existing peer review scores, which itself exhibits documented biases and inconsistencies\citep{smith2006peer}. The question of what LLMs \textit{should} align to when human judgments are themselves imperfect remains philosophically and practically open. We evaluate open-weight models exclusively and the behavior of proprietary systems, which may employ different training methods and safety interventions, could differ substantially. Finally, our post-publication metrics ($C_5$, $\text{Disruptive Index}$, $\text{Novelty \& Conventionality}$) are themselves imperfect proxies for scientific value, subject to gaming, field-specific dynamics, and temporal variation. Our study provides the most comprehensive evaluation of LLMs as peer reviewers, connecting model behavior to post-publication outcomes through validated science of science metrics.

\section*{Acknowledgement}
The experiments involved in the study were run on Google Cloud, and Lambda Labs. The computing resources for this work is supported in part by the Google Cloud Research Credits Grant 331845891, and Lambda Labs Credits through the support program D1: CSC-SUPPORT-CDFF-2025-3-31. Additionally, a small portion of the ablations in this research was supported in part through the computational resources and staff contributions provided for the Quest high performance computing facility at Northwestern University which is jointly supported by the Office of the Provost, the Office for Research, and Northwestern University Information Technology.
\bibliographystyle{unsrtnat}
\bibliography{references}

\newpage 

\appendix
\section{Prompt used for the Zero-shot peer-review}
\label{zero_shot_prompts}

\vspace{-0.5cm}
\begin{tcolorbox}[
    colback=bgray,
    colframe=borderblue,
    boxrule=1.5pt,
    arc=5mm, 
    width=15cm,
    boxsep=0mm,
    breakable,
    title=\textbf{Zero-Shot Peer Review Prompt},
    colbacktitle=titlebg,
    coltitle=white,
    fonttitle=\bfseries\sffamily,
    attach boxed title to top center={yshift=-2.5mm, yshifttext=-1mm},
    boxed title style={
        arc=3mm,
        boxrule=0.5pt,
    }
]

\tcbset{
    enhanced,
    colback=white,
    arc=3mm,
    boxrule=0.5pt,
    colframe=black!15, 
    fonttitle=\bfseries,
    left=5mm, right=5mm, top=5mm, bottom=5mm,
    boxsep=3mm,
    lefttitle=3mm,
    toptitle=2mm,
    bottomtitle=2mm
}

\begin{tcolorbox}[title=Task]
You are given Paper title, keywords and full text of a scientific paper. Your goal is to accurately analyze the entire manuscript and play the role of a peer-reviewer to evaluate the entire manuscript of the paper. You need to give a numerical score outlining your full paper review and confidence in your decision.
\end{tcolorbox}

\begin{tcolorbox}[title=Considerations, top=2mm] 
\begin{enumerate}[leftmargin=*, label=\arabic*., itemsep=2pt]
    \item You have to give a review of the idea from the range \hl{1 to 10}.
    \item You have to also give a review of the entire full text of the paper from the range \hl{1 to 10}.
    \item Your rating of the paper's idea and full text must include a confidence on the range of \hl{1 to 10}.
    \item You will be given papers across different scientific fields so be adaptable when reviewing the idea.
    \item Carefully evaluate the full content of the paper and donot jump to quick conclusions.
    \item \textbf{DO NOT} hallucinate and produce new information.
\end{enumerate}
\end{tcolorbox}

\begin{tcolorbox}[
    title=Idea Review Response Format: JSON,
    top=2mm,
    listing only, 
    listing style={style=tcblatex,
        texcsstyle=*\color{blue!70!black},
        commentstyle=\color{gray},
        stringstyle=\color{orange!80!black}
    },
    listing options={
        basicstyle=\ttfamily\small,
        upquote=true 
    }
]
{
  idea-only-review-confidence: int, \\
  idea-only-review-content: str, \\
  idea-only-review-rating: int
}
\end{tcolorbox}

\begin{tcolorbox}[
    title=Full Text Review Response Format: JSON,
    top=2mm,
    listing only, 
    listing style={style=tcblatex,
        texcsstyle=*\color{blue!70!black},
        commentstyle=\color{gray},
        stringstyle=\color{orange!80!black}
    },
    listing options={
        basicstyle=\ttfamily\small,
        upquote=true 
    }
]
{
  full-text-review-confidence: int, \\
  full-text-review-content: str, \\
  full-text-review-rating: int
}
\end{tcolorbox}

\begin{tcolorbox}[
    title=Input Text Structure,
    top=2mm,
    listing only,
    listing options={basicstyle=\ttfamily\small}
]
**Paper Title:**
TITLE

**Keywords:**
KEYWORDS

**Paper Full Text:**
FULL-TEXT
\end{tcolorbox}
\end{tcolorbox}

\newpage

\section{Prompt used for the Zero-shot peer-review of ideas presented in a paper}
\label{zero_shot_prompts_idea_review}

\vspace{-0.5cm}
\begin{tcolorbox}[
    colback=bgray,
    colframe=borderblue,
    boxrule=1.5pt,
    arc=5mm, 
    width=15cm,
    boxsep=0mm,
    breakable,
    title=\textbf{Zero-Shot Idea Review Prompt},
    colbacktitle=titlebg,
    coltitle=white,
    fonttitle=\bfseries\sffamily,
    attach boxed title to top center={yshift=-2.5mm, yshifttext=-1mm},
    boxed title style={
        arc=3mm,
        boxrule=0.5pt,
    }
]

\tcbset{
    enhanced,
    colback=white,
    arc=3mm,
    boxrule=0.5pt,
    colframe=black!15, 
    fonttitle=\bfseries,
    left=5mm, right=5mm, top=5mm, bottom=5mm,
    boxsep=3mm,
    lefttitle=3mm,
    toptitle=2mm,
    bottomtitle=2mm
}

\begin{tcolorbox}[title=Task]
You are given Paper title, abstract and keywords of a scientific paper. Your goal is to accurately analyze the entire manuscript and play the role of a peer-reviewer to evaluate the ideas presented in the paper. You need to give a numerical score outlining your review and confidence in your decision.
\end{tcolorbox}

\begin{tcolorbox}[title=Considerations, top=2mm] 
\begin{enumerate}[leftmargin=*, label=\arabic*., itemsep=2pt]
    \item  You have to give a review of the idea from the range 1 to 10.
    \item Your rating of the paper's idea must include a confidence on the range of 1 to 10.
    \item You will be given papers across different scientific fields so be adaptable when reviewing the idea.
    \item DONOT hallucinate and produce new information.
\end{enumerate}
\end{tcolorbox}

\begin{tcolorbox}[
    title=Idea Review Response Format: JSON,
    top=2mm,
    listing only, 
    listing style={style=tcblatex,
        texcsstyle=*\color{blue!70!black},
        commentstyle=\color{gray},
        stringstyle=\color{orange!80!black}
    },
    listing options={
        basicstyle=\ttfamily\small,
        upquote=true 
    }
]
{
  idea-only-review-confidence: int, \\
  idea-only-review-content: str, \\
  idea-only-review-rating: int
}
\end{tcolorbox}

\begin{tcolorbox}[
    title=Input Text Structure,
    top=2mm,
    listing only,
    listing options={basicstyle=\ttfamily\small}
]
**Paper Title:**
TITLE

**Paper Abstract:**
```plaintext
ABSTRACT
```

**Keywords:**
KEYWORDS
\end{tcolorbox}
\end{tcolorbox}

\newpage

\section{Prompt used for the Ablations on post publication outcomes dataset}
\label{zero_shot_ablations_prompts}

\vspace{-0.85cm}
\begin{tcolorbox}[
    colback=bgray,
    colframe=borderblue,
    boxrule=1.5pt,
    arc=5mm, 
    width=15cm,
    boxsep=0mm,
    breakable,
    title=\textbf{Ablations Zero-Shot Peer Review Prompt},
    colbacktitle=titlebg,
    coltitle=white,
    fonttitle=\bfseries\sffamily,
    attach boxed title to top center={yshift=-2.5mm, yshifttext=-1mm},
    boxed title style={
        arc=3mm,
        boxrule=0.5pt,
    }
]

\tcbset{
    enhanced,
    colback=white,
    arc=3mm,
    boxrule=0.5pt,
    colframe=black!15, 
    fonttitle=\bfseries,
    left=5mm, right=5mm, top=5mm, bottom=5mm,
    boxsep=3mm,
    lefttitle=3mm,
    toptitle=2mm,
    bottomtitle=2mm
}

\begin{tcolorbox}[title=Task]
You are given Paper title, keywords and full text of a scientific paper. Your goal is to accurately analyze the entire manuscript and play the role of a peer-reviewer to evaluate the entire manuscript of the paper. You need to give a numerical score outlining your full paper review and confidence in your decision.
\end{tcolorbox}

\begin{tcolorbox}[title=Considerations, top=2mm] 
\begin{enumerate}[leftmargin=*, label=\arabic*., itemsep=2pt]
    \item You have to give a review of the idea from the range \hl{1 to 10}.
    \item You have to also give a review of the entire full text of the paper from the range \hl{1 to 10}.
    \item Your rating of the paper's idea and full text must include a confidence on the range of \hl{1 to 10}.
    \item You will be given papers across different scientific fields so be adaptable when reviewing the idea.
    \item Carefully evaluate the full content of the paper and donot jump to quick conclusions.
    \item \textbf{DO NOT} hallucinate and produce new information.
\end{enumerate}
\end{tcolorbox}

\begin{tcolorbox}[title=Scoring Calibration, top=2mm] 
\begin{enumerate}[leftmargin=*, label=\arabic*., itemsep=2pt]
    \item 1-3: Fundamentally flawed
    \item 4-5: Below average, limited contribution
    \item 6-7: Solid work, useful but not exceptional
    \item 8: Exceptional - top 10\% material
    \item 9: Outstanding - top 5\% material
    \item 10: Paradigm-shifting - top 1\% material
    \item Only ~5\% of papers should be considered truly exceptional (score 8+)
    \item Most papers, even good ones, should score between 4-7
    \item Reserve scores of 9-10 for paradigm-shifting work only
\end{enumerate}
\end{tcolorbox}

\begin{tcolorbox}[
    title=Idea Review Response Format: JSON,
    top=2mm,
    listing only, 
    listing style={style=tcblatex,
        texcsstyle=*\color{blue!70!black},
        commentstyle=\color{gray},
        stringstyle=\color{orange!80!black}
    },
    listing options={
        basicstyle=\ttfamily\small,
        upquote=true 
    }
]
{
  idea-only-review-confidence: int, \\
  idea-only-review-content: str, \\
  idea-only-review-rating: int
}
\end{tcolorbox}

\begin{tcolorbox}[
    title=Full Text Review Response Format: JSON,
    top=2mm,
    listing only, 
    listing style={style=tcblatex,
        texcsstyle=*\color{blue!70!black},
        commentstyle=\color{gray},
        stringstyle=\color{orange!80!black}
    },
    listing options={
        basicstyle=\ttfamily\small,
        upquote=true 
    }
]
{
  full-text-review-confidence: int, \\
  full-text-review-content: str, \\
  full-text-review-rating: int
}
\end{tcolorbox}

\begin{tcolorbox}[
    title=Input Text Structure,
    top=2mm,
    listing only,
    listing options={basicstyle=\ttfamily\small}
]
**Paper Title:**
TITLE

**Keywords:**
KEYWORDS

**Paper Full Text:**
FULL-TEXT
\end{tcolorbox}

\end{tcolorbox}

\newpage

\section{LLM-reported confidence vs Prediction Bias plots for the remaining models for idea reviews}
\label{confplots-idea-yllm-ytrue}
\begin{figure}[!htb]
    \centering
    \subfloat[\centering Gemma3-27b-IT]{{\includegraphics[width=0.3\textwidth]{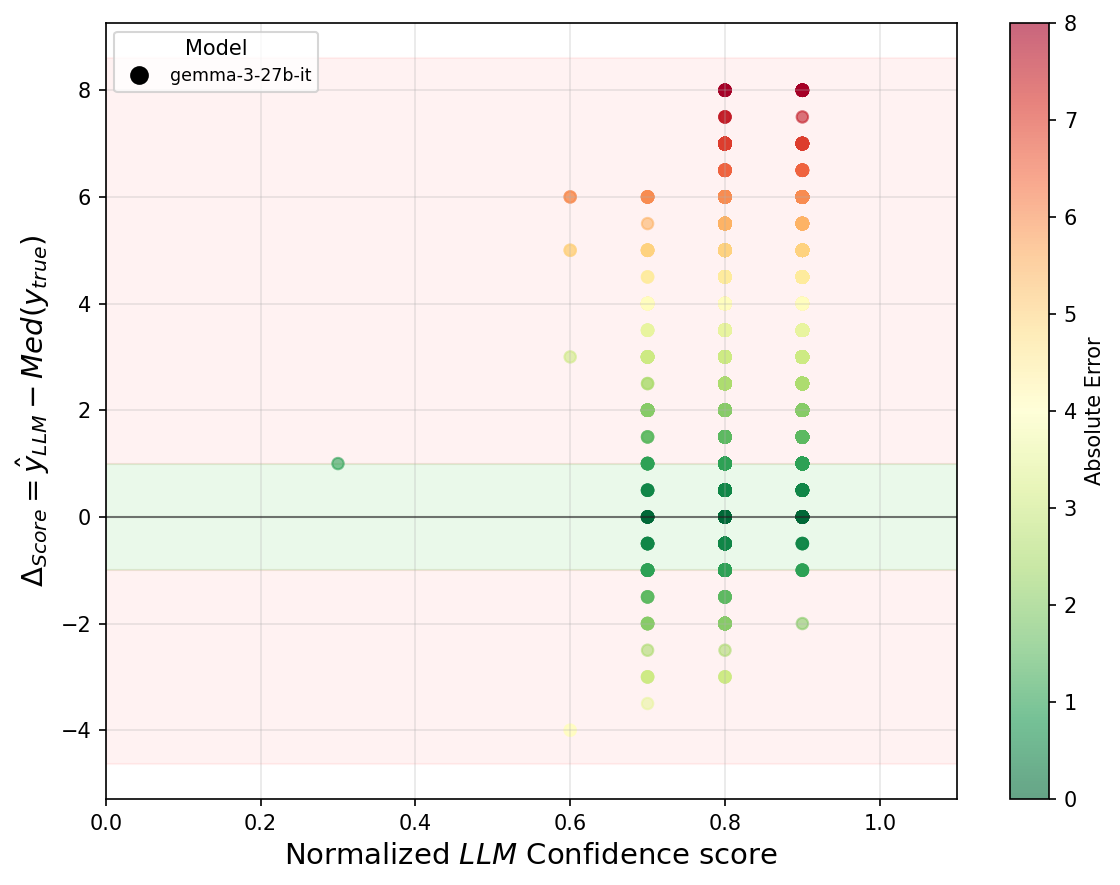} }}%
    \qquad
    \subfloat[\centering Llama-3.1-Tulu-3-70B]{{\includegraphics[width=0.3\textwidth]{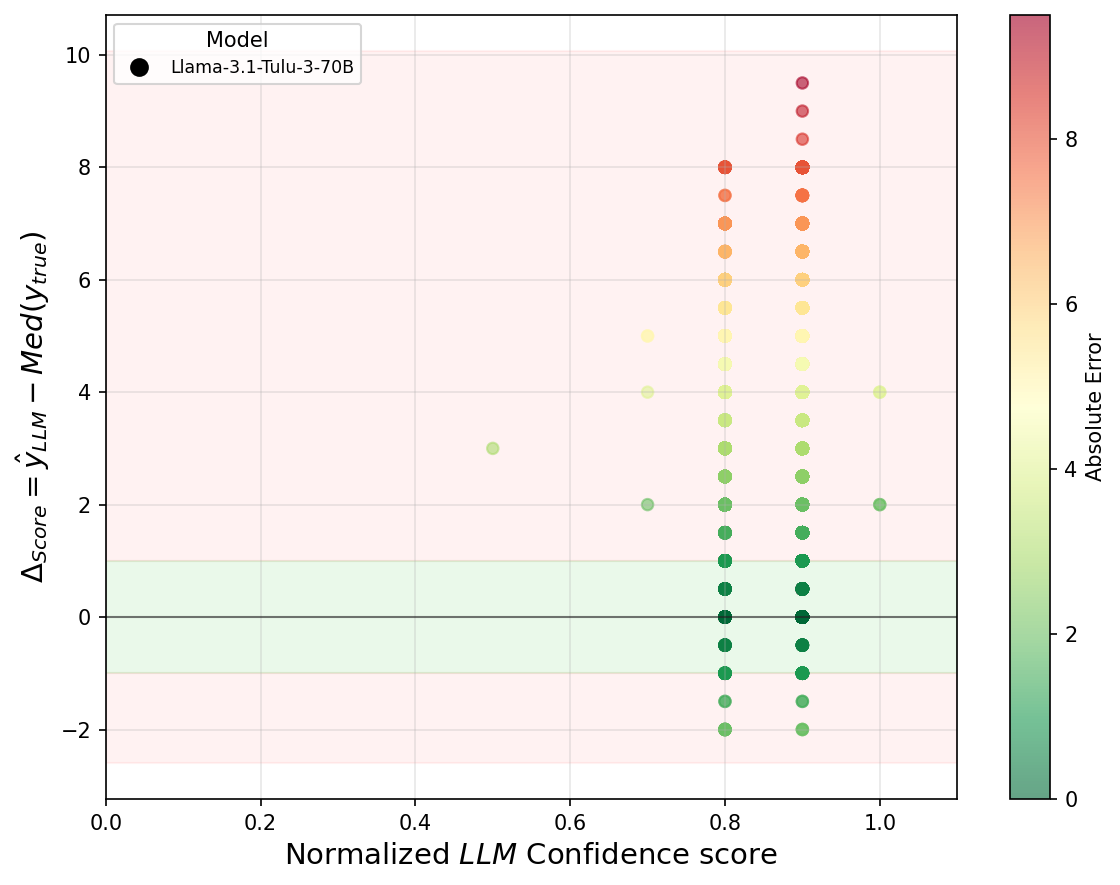} }}%
    \subfloat[\centering Llama-3.3-Nemotron-Super-49B-v1.5]{{\includegraphics[width=0.3\textwidth]{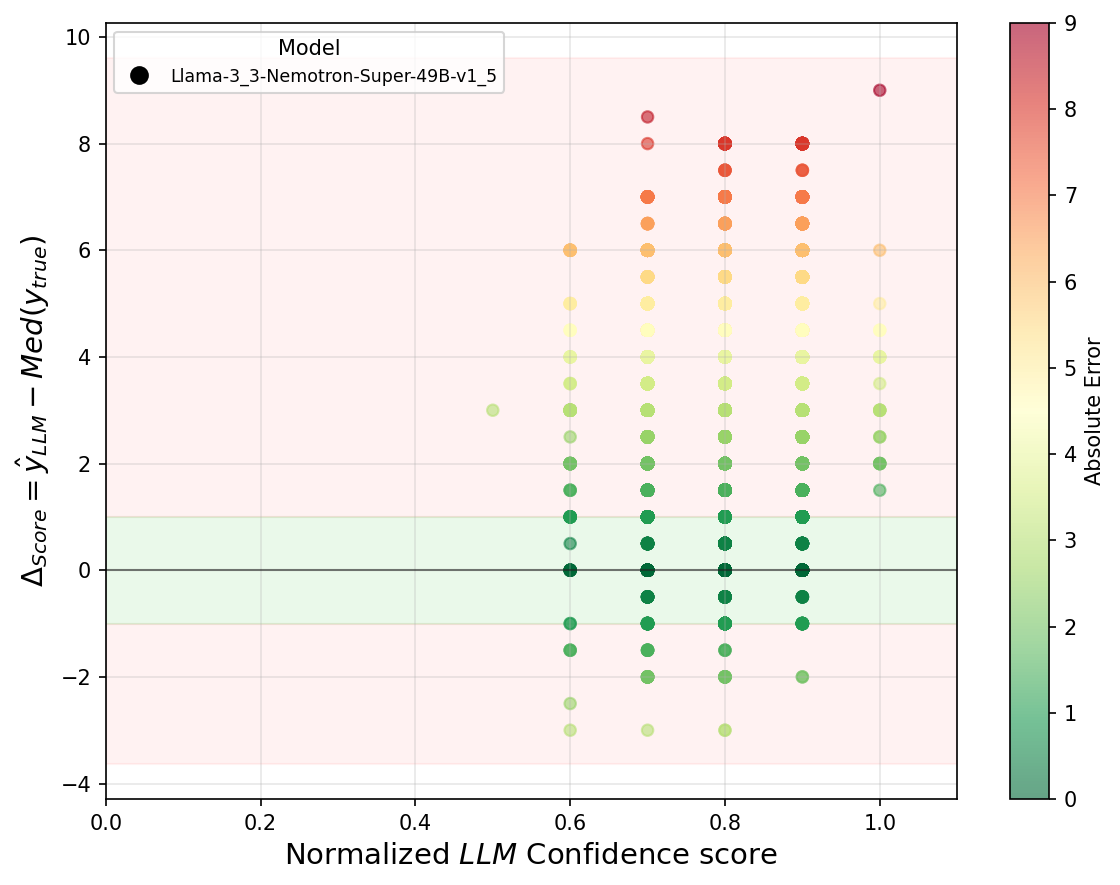} }}%
    \caption{Confidence plot when LLMs review ideas in a given paper using abstract}
    \label{fig:LMRSD_exp1_conf_plot_remaining}
\end{figure}

\section{Venue composition of the papers}
\label{venue_composition}

\begin{table}[!htb]
  \caption{Venue composition of the papers}
  \label{tab:venue_composition}
  \centering
  \begin{small}
  \begin{tabular}{l r r}
    \toprule
    \textbf{Venue} & \textbf{Year} & {\textbf{\# Papers}} \\
    \midrule
    ICLR.cc/2023/Conference                 & 2023 & 4,903 \\
    ICLR.cc/2022/Conference                 & 2022 & 3,369 \\
    ICLR.cc/2021/Conference                 & 2021 & 2,975 \\
    NeurIPS.cc/2022/Conference              & 2022 & 2,816 \\
    NeurIPS.cc/2021/Conference              & 2021 & 2,754 \\
    ICLR.cc/2020/Conference                 & 2020 & 1,732 \\
    ICLR.cc/2019/Conference                 & 2019 & 1,555 \\
    ICLR.cc/2018/Conference                 & 2018 &   980 \\
    ICLR.cc/2017/Conference                 & 2017 &   497 \\
    ICLR.cc/2018/Workshop                   & 2018 &   264 \\
    auai.org/UAI/2022/Conference            & 2022 &   229 \\
    robot-learning.org/CoRL/2023/Conference & 2023 &   197 \\
    robot-learning.org/CoRL/2022/Conference & 2022 &   197 \\
    NeurIPS.cc/2022/Track/Datasets          & 2022 &   163 \\
    robot-learning.org/CoRL/2021/Conference & 2021 &   153 \\
    NeurIPS.cc/2021/Track/Datasets          & 2021 &   144 \\
    MIDL.io/2020/Conference                 & 2020 &   135 \\
    ICLR.cc/2017/Workshop                   & 2017 &   121 \\
    MIDL.io/2019/Conference                 & 2019 &   117 \\
    MIDL.io/2023/Conference                 & 2023 &   112 \\
    \bottomrule
  \end{tabular}
  \end{small}
\end{table}

\newpage

\section{LLM-reported confidence vs Prediction Bias plots for the remaining models}
\label{confplots-yllm-ytrue}
\begin{figure}[!htb]
    \centering
    \subfloat[\centering Gemma3-27b-IT]{{\includegraphics[width=0.3\linewidth]{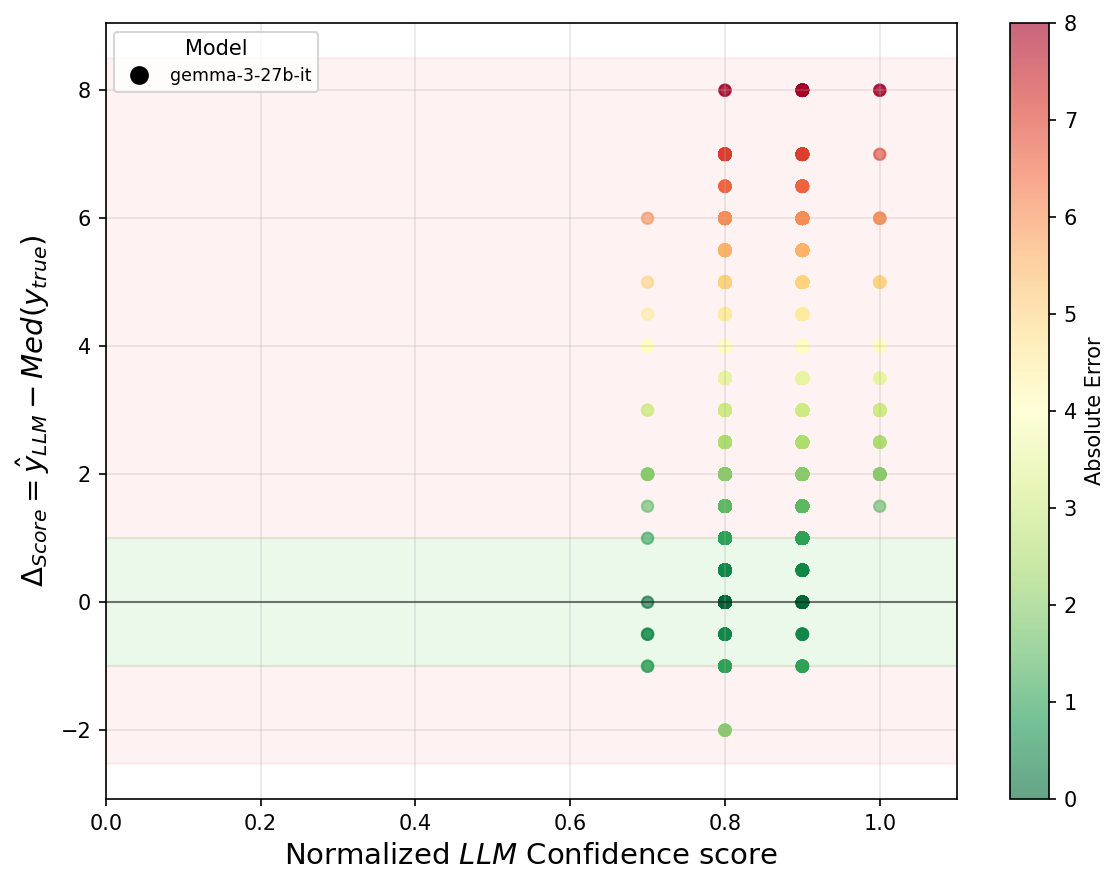} }}%
    \qquad
    \subfloat[\centering Llama-3.1-Tulu-3-70B]{{\includegraphics[width=0.3\linewidth]{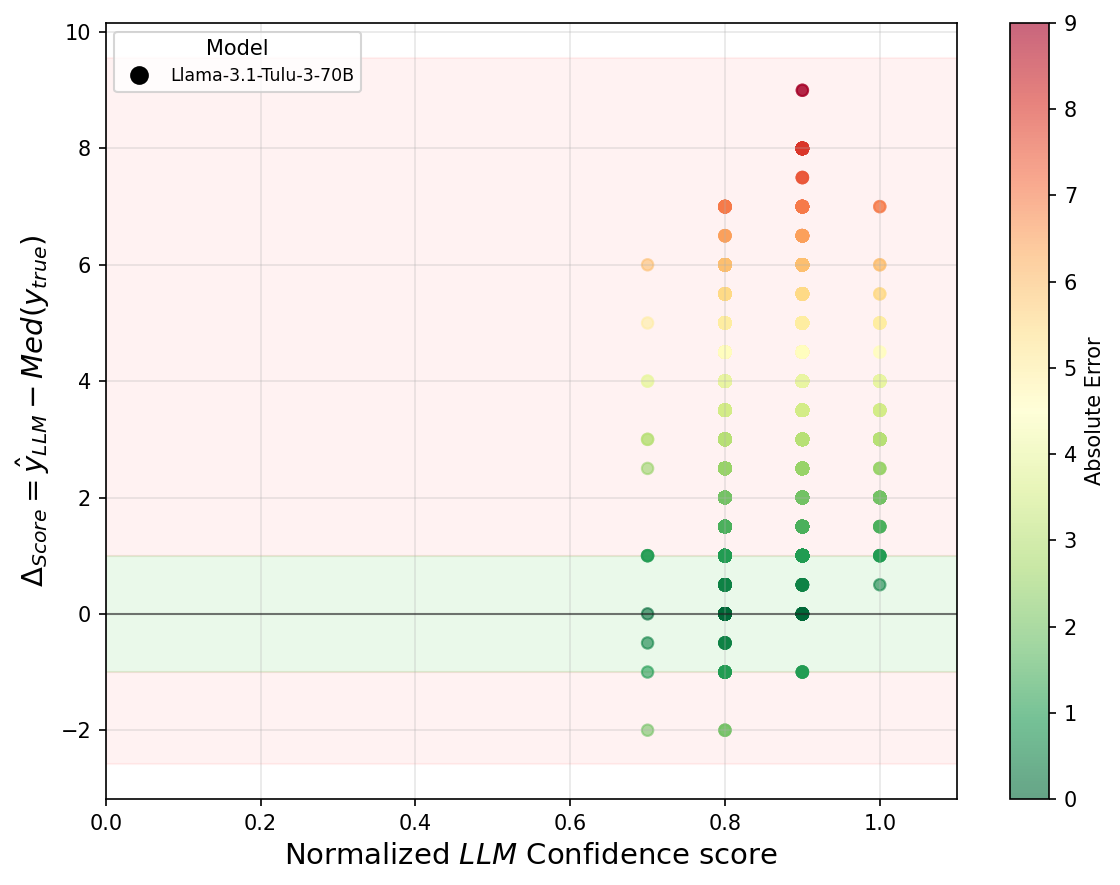} }}%
    \subfloat[\centering Llama-3.3-Nemotron-Super-49B-v1.5]{{\includegraphics[width=0.3\linewidth]{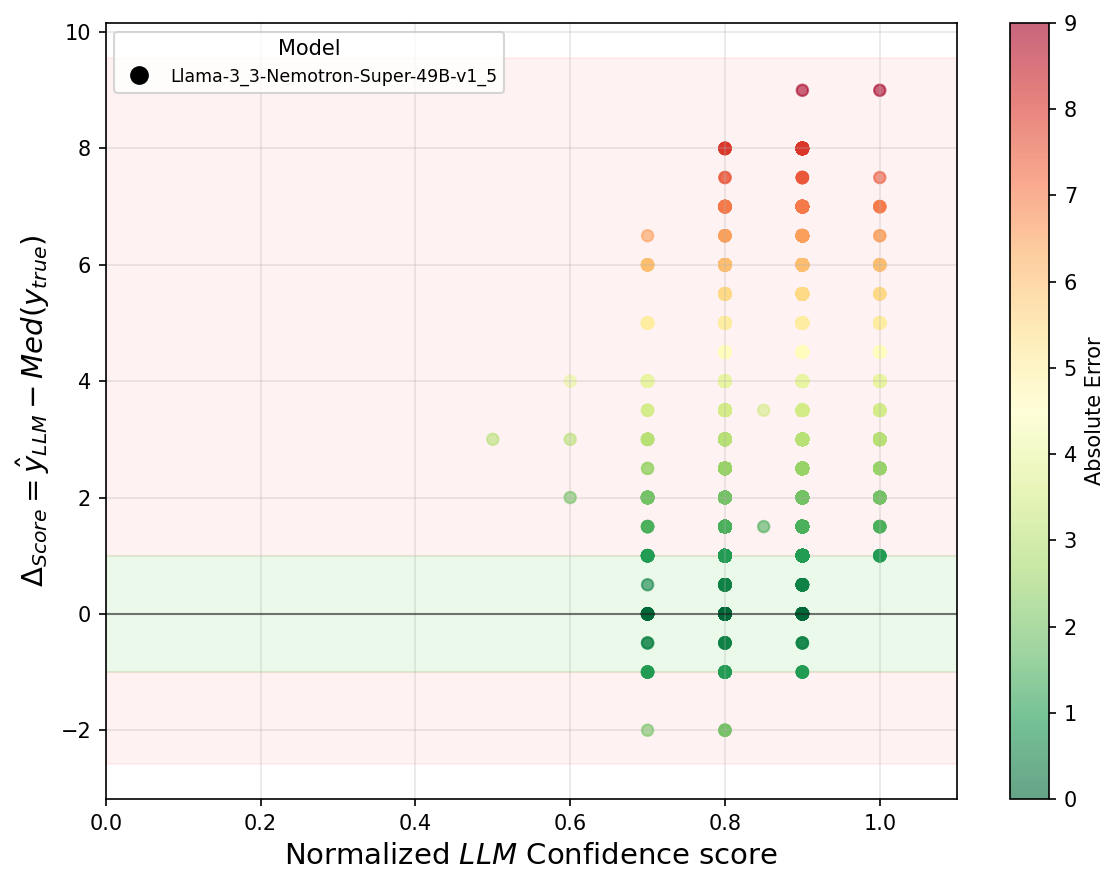} }}%
    \qquad
    \subfloat[\centering Llama-3.3-70B-Instruct]{{\includegraphics[width=0.35\linewidth]{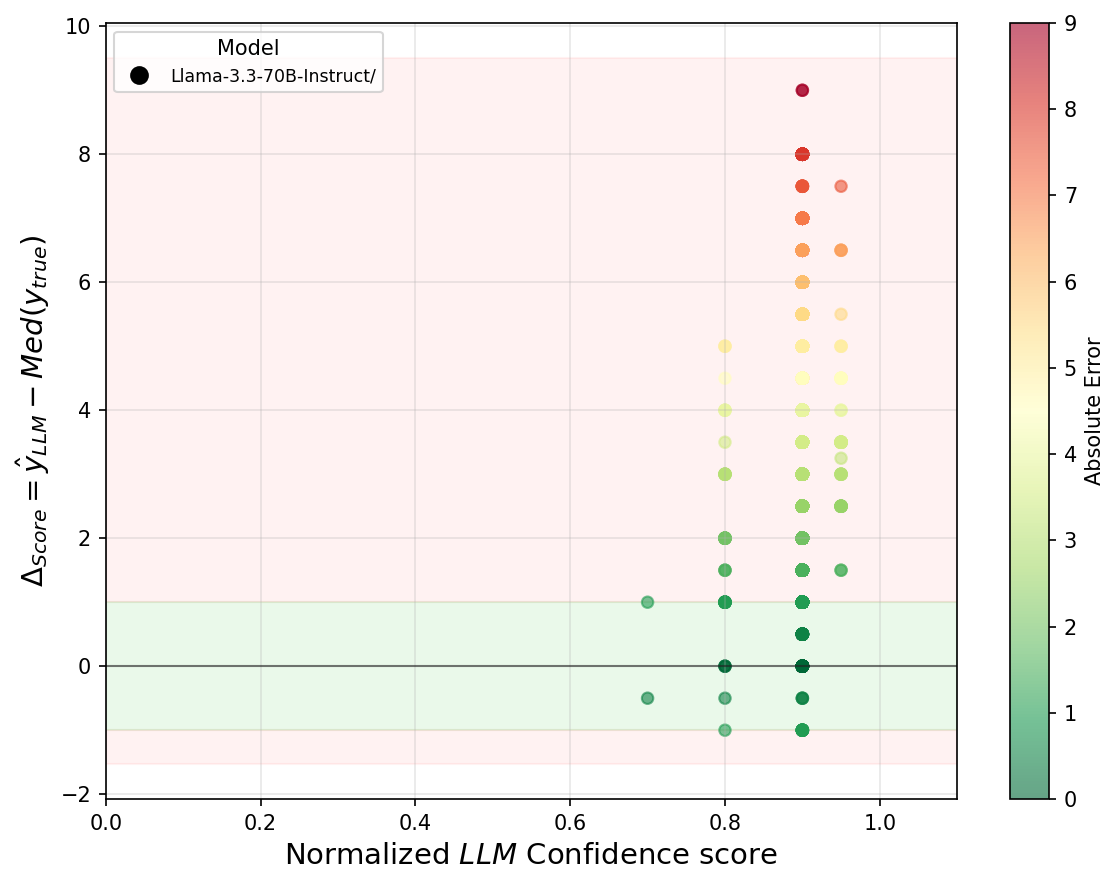} }}%
    \qquad
    \subfloat[\centering Qwen3-32B]{{\includegraphics[width=0.35\linewidth]{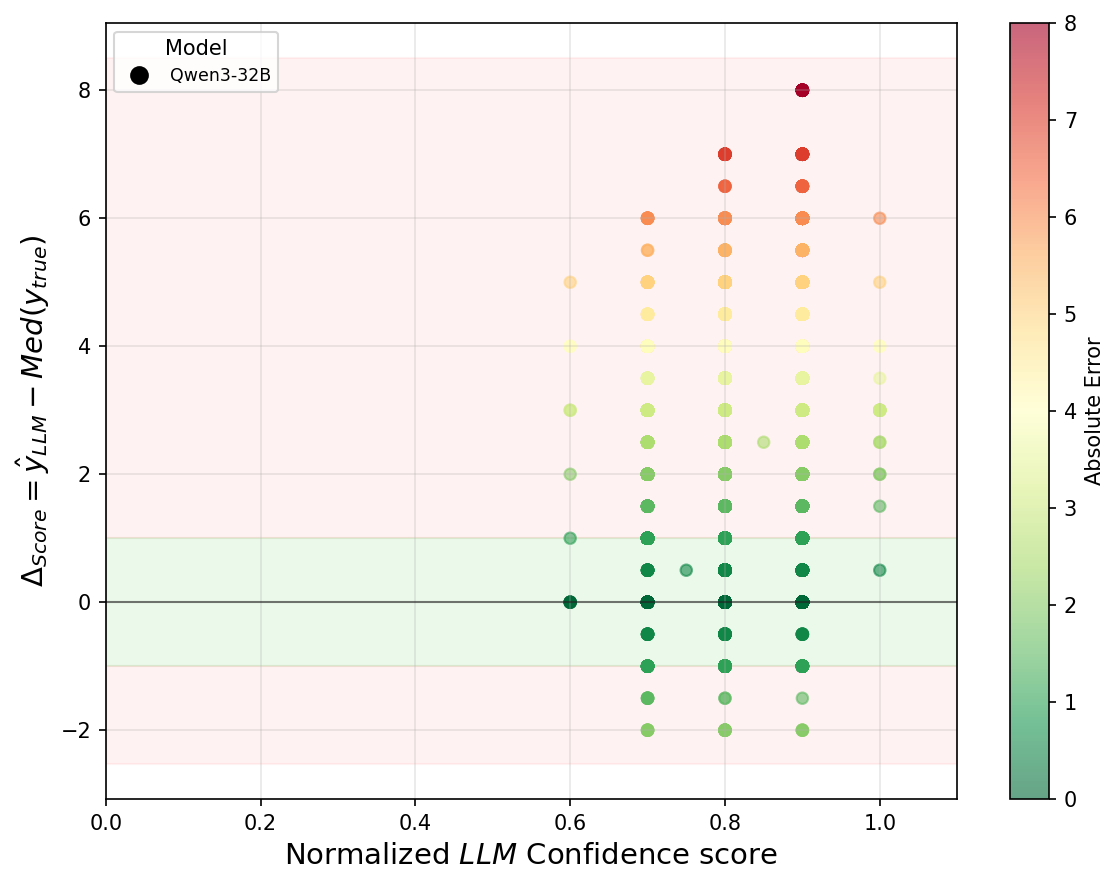} }}%
    \qquad
    \subfloat[\centering Qwen3-Next-80B-A3B-Thinking]{{\includegraphics[width=0.35\linewidth]{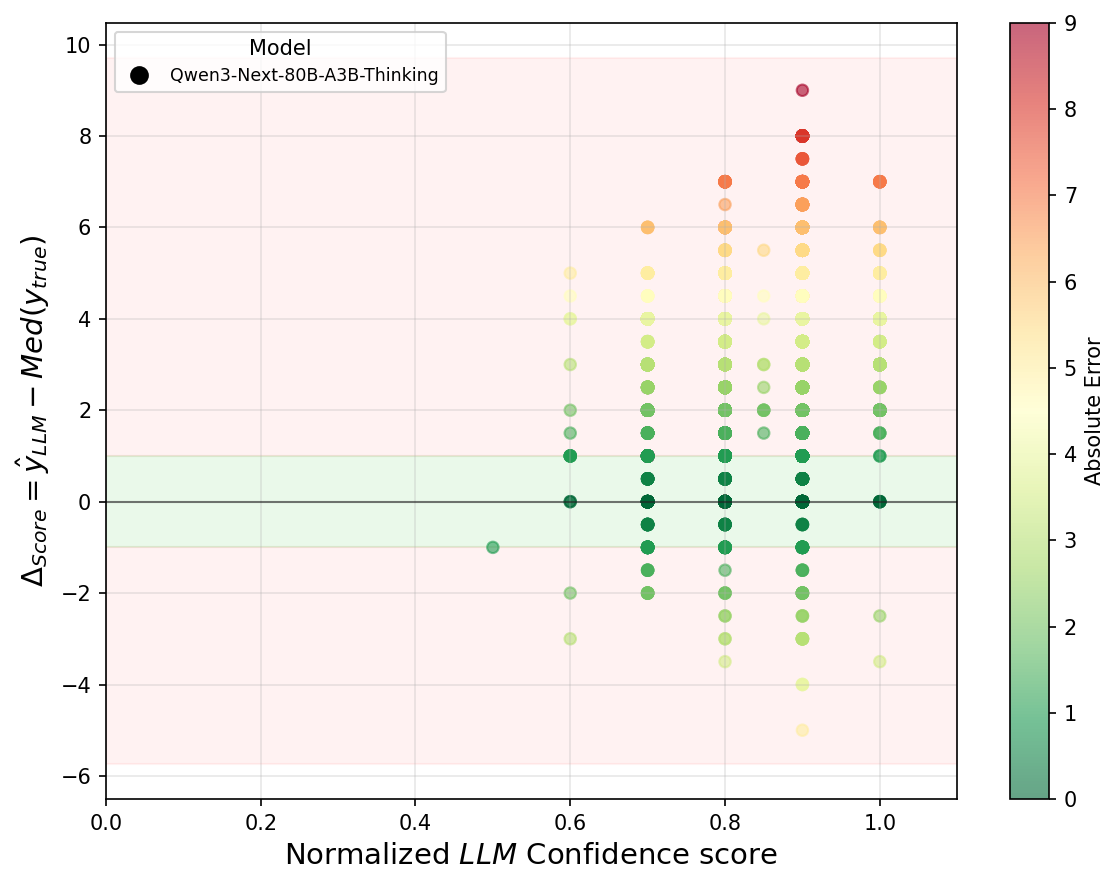} }}%
    \qquad
    \subfloat[\centering gpt-oss-20b]{{\includegraphics[width=0.35\linewidth]{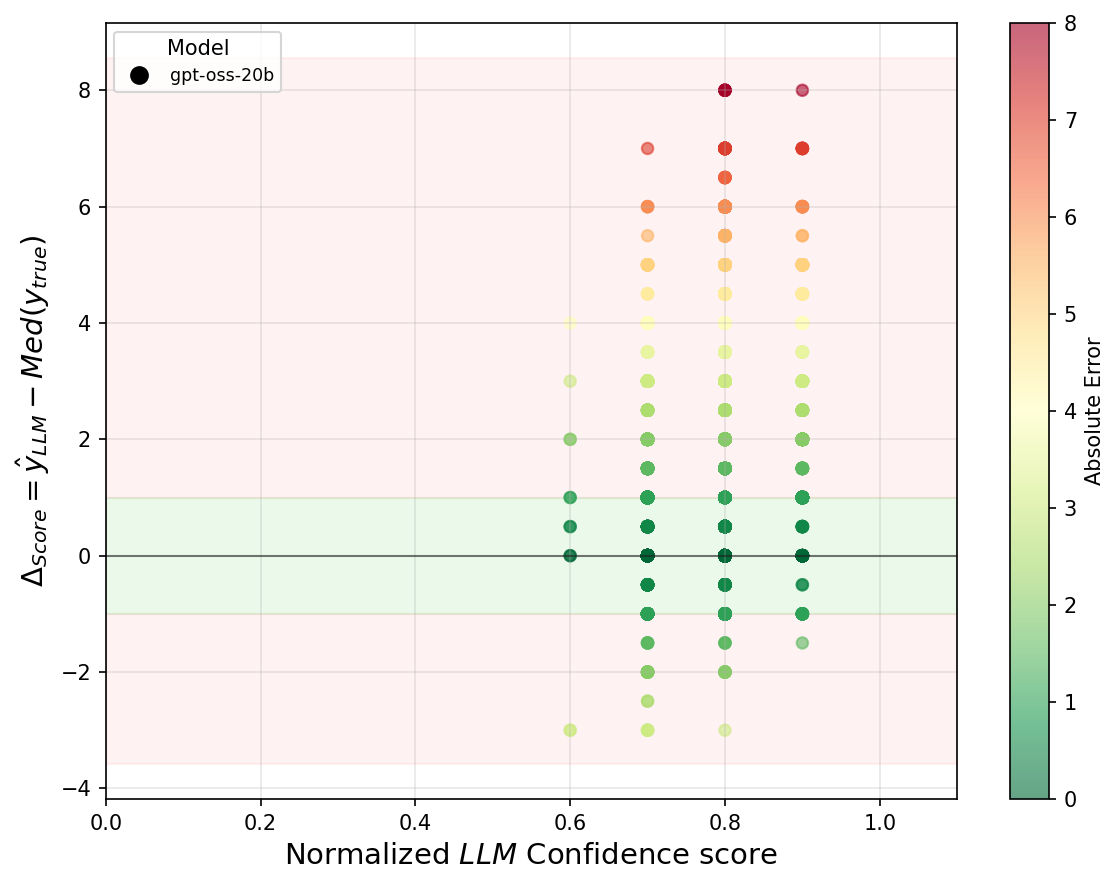} }}%

    \caption{LLM-reported Confidence vs Prediction Bias when reviewing complete content presented in a scientific article using the full-text}
    \label{fig:LMRSD_exp2_conf_plot_remaining}
\end{figure}

\section{Demonstration of Agentic Peer-review}
\label{agentic_peer_review}
\begin{figure}[!htb]
    \centering
    {\includegraphics[width=0.95\linewidth]{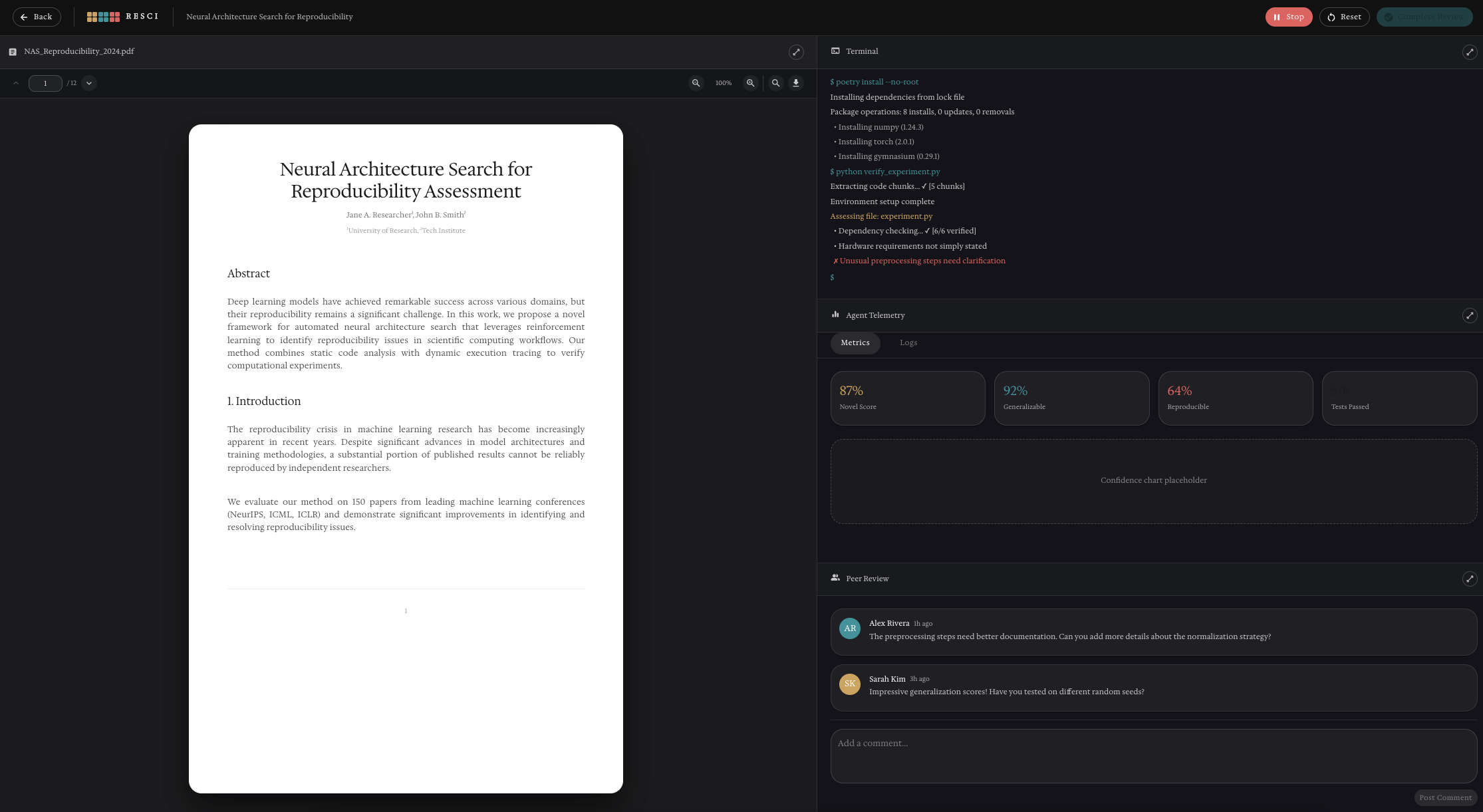} }
    \caption{Agentic Peer-review sandbox}
    \label{fig:lmrsd_agentic_peer_review}%
\end{figure}

Figure~\ref{fig:lmrsd_agentic_peer_review} demonstrates a mock interface that could potentially be useful in an agentic peer-review system since it addresses the limitations identified in our main experiments. Unlike the text-only zero-shot approaches evaluated in our prior experiments, this sandbox like interface equips any LLM with computational tools including code execution capabilities, dependency checking, and automated reproducibility assessment. 

The interface shows a multi-phase review process where the agent first analyzes the manuscript PDF, executes provided code to verify claims, and generates structured feedback. The telemetry panel displays quantitative metrics (Novelty, Generalizability, Reproducibility) computed through tool-augmented analysis rather than pure text. This approach directly responds to our finding that current LLMs exhibit systematic overconfidence and poor calibration when relying solely on manuscript text (§\ref{RQ2}). Agentic systems using similar architectures will offer a pathway toward more reliable scientific evaluation.

\end{document}